\documentclass[twocolumn, aps, prl, superscriptaddress, longbibliography]{revtex4-1}
\usepackage[latin9]{inputenc}
\usepackage{color}
\usepackage{amsmath}
\usepackage{amssymb}
\usepackage{graphicx}
\usepackage[unicode=true,
 bookmarks=false,
 breaklinks=false,pdfborder={0 0 1},backref=false,colorlinks=true,urlcolor=black]
 {hyperref}
\hypersetup{
 colorlinks, citecolor=blue}

\makeatletter

\usepackage{amsfonts}\usepackage{amsthm}
\usepackage{graphicx}
\usepackage{xcolor}
\usepackage{amsmath}

\newcommand{\ket}[1]{|{#1}\rangle}

\renewcommand\[{\begin{equation}}
\renewcommand\]{\end{equation}}

\theoremstyle{plain}

\makeatother

\begin{document}

\title{Scale-Invariant Continuous Entanglement Renormalization of a Chern
Insulator}

\author{Su-Kuan Chu}
\affiliation{Joint Quantum Institute,
NIST/University of Maryland, College Park, Maryland 20742, USA}
\affiliation{Joint Center for Quantum Information and Computer Science,
NIST/University of Maryland, College Park, Maryland 20742, USA}

\author{Guanyu Zhu}
\affiliation{Joint Quantum Institute,
NIST/University of Maryland, College Park, Maryland 20742, USA}

\author{James R. Garrison}
\affiliation{Joint Quantum Institute,
NIST/University of Maryland, College Park, Maryland 20742, USA}
\affiliation{Joint Center for Quantum Information and Computer Science,
NIST/University of Maryland, College Park, Maryland 20742, USA}

\author{Zachary Eldredge}
\affiliation{Joint Quantum Institute,
NIST/University of Maryland, College Park, Maryland 20742, USA}
\affiliation{Joint Center for Quantum Information and Computer Science,
NIST/University of Maryland, College Park, Maryland 20742, USA}

\author{Ana~Vald\'es Curiel}
\affiliation{Joint Quantum Institute,
NIST/University of Maryland, College Park, Maryland 20742, USA}

\author{Przemyslaw Bienias}
\affiliation{Joint Quantum Institute,
NIST/University of Maryland, College Park, Maryland 20742, USA}

\author{I. B. Spielman}
\affiliation{Joint Quantum Institute,
NIST/University of Maryland, College Park, Maryland 20742, USA}

\author{Alexey V. Gorshkov}
\affiliation{Joint Quantum Institute,
NIST/University of Maryland, College Park, Maryland 20742, USA}
\affiliation{Joint Center for Quantum Information and Computer Science,
NIST/University of Maryland, College Park, Maryland 20742, USA}
\begin{abstract}
The multi-scale entanglement renormalization ansatz (MERA) postulates the existence of quantum circuits that renormalize entanglement in real space at different length scales. Chern insulators, however, cannot have scale-invariant discrete MERA circuits with finite bond dimension. In this Letter, we show that the continuous MERA (cMERA), a modified version of MERA adapted for field theories, possesses a fixed point wavefunction with nonzero Chern number. Additionally, it is well known that reversed MERA circuits can be used to prepare quantum states efficiently in time that scales logarithmically with the size of the system. However, state preparation via MERA typically requires the advent of a full-fledged universal quantum computer. In this Letter, we demonstrate that our cMERA circuit can potentially be realized in existing analog quantum computers, i.e., an ultracold atomic Fermi gas in an optical lattice with light-induced spin-orbit coupling.
\end{abstract}
\maketitle
A quantum many-body system has a Hilbert space whose dimension grows exponentially with system size, making
exact diagonalization of its Hamiltonian impractical. Fortunately, tensor networks \cite{Orus2014,Bridgeman2016}
are capable of efficiently representing
the ground states of many systems with local interactions \cite{Vidal2006,Verstraete2006,Verstraete2006a,Hastings2007,Vidal2007,Wolf2008}.
Another powerful tool in many-body physics is the renormalization
group (RG) \cite{Wilson1974,Wilson1975}, which uses the fact
that the description of a physical system can vary at different length
scales, forming a hierarchical structure. The RG provides a systematic prescription to transform an exact microscopic description to an effective coarse-grained description. Applications of RG range
from critical phenomena in condensed matter to the electroweak interaction
in high-energy physics \cite{Zinn-Justin2007}.

One approach which combines tensor networks and renormalization group is called the multi-scale entanglement renormalization ansatz (MERA)
\cite{Vidal2006,Vidal2007}. MERA proposes a quantum circuit
acting on a state which is initially entangled at many length scales.
The two elementary building-block tensors of the MERA, isometries
and disentanglers, are discrete unitary gates which physically implement RG in
real space by successively removing entanglement at progressively larger length scales. Interestingly, since
the circuit depth only increases logarithmically with the system size,
a reversed MERA circuit can efficiently prepare a state
with finer entanglement structure from a weakly-entangled initial
state. In practice, MERAs are most convenient when the disentanglers
and isometries are independent of the length scale \cite{Aguado2008,Pfeifer2009,Konig2009,Konig2010,Singh2013,Evenbly2016,Haegeman2017}.
The state that is left unchanged in the thermodynamic limit by these
scale-invariant unitaries is termed a fixed-point wavefunction.

Experimentally, a reversed MERA circuit might be used to prepare exotic states,
such as chiral topological states, which include integer quantum Hall
states and certain fractional quantum Hall states \cite{Hansson2016,Wen2016a}.
Some fractional quantum Hall systems are believed to feature anyons
useful for topological quantum computation \cite{Nayak2008}. Due to their great theoretical interest,
it would be useful to be able to study these systems under highly
controlled settings, such as in ultracold atomic gases. However, to
create a chiral topological state in the lab, we must not only engineer
the parent Hamiltonian, but also cool the system down
to the ground state. The latter is usually hard experimentally for
topological states due to their long-range entanglement \cite{Bravyi2006}.
A reversed MERA circuit can possibly resolve this issue by directly generating the target
chiral topological state from another state that is easier to obtain
by cooling. 

Here, as a first step towards finding a MERA for a fractional quantum Hall state, we instead search
for a MERA whose fixed-point wavefunction describes an (integer) Chern insulator.
A Chern insulator is an integer quantum Hall state on a lattice and is therefore a simpler system than the fractional quantum Hall state. However, there are no-go
theorems stating that a MERA cannot have a Chern insulator ground state as its fixed-point wavefunction
\cite{Barthel2010,Dubail2015,Wahl2013,Li2017}. Since
conventional MERA only contains strictly local interactions, adding
quasi-local interactions with quickly decaying tails could possibly
circumvent the no-go theorems. A modified formalism of MERA adapted
for field theories called continuous MERA (cMERA) \cite{Haegeman2013}
can include such quasi-local interactions \cite{Hu2017}. In contrast
to the MERA paradigm, in which the renormalization circuit consists
of discrete unitary gates, cMERA treats the circuit time, which corresponds to the length scale, as a continuous variable and generates continuous entanglement
renormalization using a Hermitian Hamiltonian. 

In this Letter, we show that a type of Chern insulator wavefunction
can be generated by a scale-invariant cMERA circuit. The Chern insulator
model we consider is the Bernevig-Hughes-Zhang model in the continuum limit \cite{Bernevig2006a}.
In addition, we propose a possible experimental realization of the
cMERA circuit with neutral $^{171}\mathrm{Yb}$ atoms in an optical
lattice by introducing spin-orbit coupling.

Our work complements and can be contrasted with Refs.\ \cite{Wen2016,Swingle2016}.
While Ref.\ \cite{Wen2016} previously developed a cMERA for the continuous Chern insulator
model mentioned above, our work uses a scale-invariant disentangler.
Other prior work in Ref.\ \cite{Swingle2016} presented a scale-invariant entanglement
renormalization for a two-band Chern insulator model. 
While Ref.\ \cite{Swingle2016} makes
use of the lattice structure and quasi-adiabatic paths between a series of gapped Hamiltonians, our cMERA
approach allows smooth time evolution and emphasizes the continuum
physics. Another difference is that the RG evolution in Ref.\ \cite{Swingle2016}
involves interactions decaying with distance faster than any power-law
function but slower than an exponential, whereas our cMERA only needs
an exponentially decaying interaction. Other known
methods for representing chiral topological states include artificial
neural network quantum states \cite{Huang2017,Kaubruegger2017,Glasser2017},
projected entangled pair states \cite{Wahl2013,Wahl2014,Poilblanc2015,Poilblanc2016},
matrix product states \cite{Zaletel2012a}, and polynomial-depth unitary
circuits \cite{Schmoll2017a}.

\textit{Review of cMERA.\textemdash{}}Within the framework of conventional MERA \cite{Vidal2006,Vidal2007},
disentanglers $V_{u}$ and isometries $W_{u}$ are strictly local
discrete unitary operators employed to renormalize entanglement at
layer $u\in\mathbb{Z}^{+}$. In cMERA \cite{Haegeman2013},
we simply replace them by continuous unitary transformations, which
are infinitesimally generated by self-adjoint operators $K(u)$ and
$L$: $V_{u}\rightarrow e^{-iK(u)\text{d}u}$, $W_{u}\rightarrow e^{-iL\text{d}u}$.
The notation $\text{d}u$ denotes an infinitesimal RG step, and $u\in(-\infty,0]$.
When the continuous variable $u$ approaches zero, the system is said to be
at the ultraviolet (UV) length scale, possessing both short-range
and long-range entanglement. As $u\rightarrow-\infty$, the system
flows to the infrared (IR) length scale, where short-range entanglement
is removed and nearly all degrees of freedom are disentangled from
each other. Note that the generator of disentangler $K(u)$ can in
general depend on scale $u$. A cMERA is called scale-invariant if $K(u)$ is independent of $u$.

To emulate the coarse-graining behavior of isometries in conventional
lattice MERA, $L$ is chosen to be the scaling transformation in field
theory. For example, for a single fermion field $\psi(\mathbf{x})$ in $d$ spatial
dimensions, we pick $L=-\frac{i}{2}\int\left(\psi^{\dagger}(\mathbf{x})\,\mathbf{x}\cdot\nabla\psi(\mathbf{x})-\mathbf{x}\cdot\nabla\psi^{\dagger}(\mathbf{x})\,\psi(\mathbf{x})\right)\mathrm{d}^{d}\mathbf{x}$
\cite{Haegeman2013,Wen2016}; thereby, fermionic operators $\psi(\mathbf{x})$
in real space and $\psi(\mathbf{k})$ in momentum space satisfy
the following scaling transformations: $e^{-iuL}\psi(\mathbf{x})e^{iuL}=e^{\frac{d}{2}u}\psi(e^{u}\mathbf{x})$,
$e^{-iuL}\psi(\mathbf{k})e^{iuL}=e^{-\frac{d}{2}u}\psi(e^{-u}\mathbf{k})$.
One can check that the anti-commutation relations $\{\psi(\mathbf{x}),\psi^{\dagger}(\mathbf{x}')\}=\delta(\mathbf{x}-\mathbf{x}')$
in real space and $\{\psi(\mathbf{k}),\psi^{\dagger}(\mathbf{k}')\}=\delta(\mathbf{k}-\mathbf{k}')$
in momentum space are preserved under the scaling transformation.
We will sometimes abuse the terminology to call $K(u)$ and $L$ themselves
the disentangler and the isometry rather than the verbose generator
of disentangler and generator of isometry. 

The renormalized wavefunction is governed by the Schr\"odinger equation,
\begin{equation}
i\frac{\partial}{\partial u}\ket{\Psi^{S}(u)}=[K(u)+L]\ket{\Psi^{S}(u)},\label{eq:schrodingerpic}
\end{equation}
where the superscript $S$ denotes the Schr\"odinger picture. In this Letter, we will focus on the interaction picture which provides a more
convenient way to look at continuous entanglement renormalization. We treat $L$
as a ``free'' Hamiltonian and $K(u)$ as an ``interaction'' Hamiltonian,
i.e., $\ket{\Psi^{I}(u)}=e^{iuL}\ket{\Psi^{S}(u)}$, where the superscript
$I$ denotes the interaction picture. Substituting this equation into
Eq.\ (\ref{eq:schrodingerpic}), we obtain 
\begin{equation}
i\frac{\partial}{\partial u}\ket{\Psi^{I}(u)}=\hat{K}(u)\ket{\Psi^{I}(u)},\label{eq:interactioneq}
\end{equation}
where $\hat{K}(u)\overset{\mathrm{{def}}}{=}e^{iuL}K(u)e^{-iuL}$
is the disentangler in the interaction picture. The renormalized wavefunction
$\ket{\Psi^{I}(u)}$ at scale $u$ can be formally written in terms
of the IR state $\ket{\Omega_\mathrm{IR}^{I}}\equiv\ket{\Psi^{I}(u\rightarrow-\infty)}$
as
\begin{equation}
\ket{\Psi^{I}(u)}=\mathcal{P}\exp\left(-i\int_{-\infty}^{u}\hat{K}(u')\mathrm{d}u'\right)\ket{\Omega_\mathrm{IR}^{I}},\label{eq:stateevo}
\end{equation}
where $\mathcal{P}$ is the path ordering operator. Unless otherwise
stated, we will only consider the interaction picture; therefore,
we will drop the superscript $I$ in the rest of this Letter.

\textit{A continuous Chern insulator model.\textemdash{}}We begin with a two-band continuous Chern insulator
model in two spatial dimensions \cite{Bernevig2006a} with Hamiltonian $H=\int \mathrm{d}^{2}\mathbf{k}\,\psi^{\dagger}(\mathbb{\mathbf{k}})[\mathbf{R}(\mathbf{k})\cdot\boldsymbol{\sigma}]\psi(\mathbf{k}),$
where $\mathbf{k}=(k_{x},\:k_{y})\in\mathbb{R}^{2}$, $\mathbf{R}(\mathbf{k})=(k_{x},\:k_{y},\:m-k^{2}),\:m>0,\:k\equiv|\mathbf{k}|=\sqrt{k_{x}^{2}+k_{y}^{2}}$,
and $\boldsymbol{\sigma}=(\sigma_{x},\:\sigma_{y},\:\sigma_{z})$
is a vector of Pauli matrices. The fermionic operator $\psi(\mathbf{k})$
is a two-component spinor $\psi(\mathbf{k})\equiv\left(\begin{array}{cc}
\psi_{1}(\mathbf{k}) & \psi_{2}(\mathbf{k})\end{array}\right)^{T}$ whose components satisfy $\{\psi_{i}^{\dagger}(\mathbf{k}),\psi_{j}(\mathbf{k}')\}=\delta_{ij}\,\delta(\mathbf{k}-\mathbf{k}')$
for $i,j\in\left\{ 1,2\right\} $.

The ground state, which has the lower band filled, is \cite{Wen2016}
\begin{align}
\ket{\Psi} & =\prod_{\mathbf{k}}\left(u_{\mathbf{k}}\psi_{2}^{\dagger}(\mathbf{k})-v_{\mathbf{k}}\psi_{1}^{\dagger}(\mathbf{k})\right)\ket{\mathrm{vac}},\label{eq:groundstate}\\
u_{\mathbf{k}} & =\frac{1}{\sqrt{N_{k}}}\left(\left(m-k^{2}\right)+\sqrt{(m-k^{2})^{2}+k^{2}}\right),\nonumber \\
v_{\mathbf{k}} & =\frac{1}{\sqrt{N_{k}}}\left(ke^{-i\theta_{k}}\right),\nonumber 
\end{align}
where $N_{k}$ is a $k$-dependent normalization factor such that
$|u_{\mathbf{k}}|^{2}+|v_{\mathbf{k}}|^{2}=1$, and the state $\ket{\mathrm{vac}}$
is the vacuum state annihilated by $\psi_{1,2}(\mathbf{k})$. The
angle $\theta_{k}$ is defined via $k_{x}=k\cos\theta_{k}$ and $k_{y}=k\sin\theta_{k}$,
i.e., it is the polar angle in momentum space. The Chern number of the bottom band of this two-band
system is $C=\frac{1}{4\pi}\int_{\mathbb{R}^2}\mathrm{d}^{2}\mathbf{k}\:\mathbf{\mathbf{n}}(\mathbf{k})\cdot\left(\frac{\partial\mathbf{n}(\mathbf{k})}{\partial k_{x}}\times\frac{\partial\mathbf{n}(\mathbf{k})}{\partial k_{y}}\right)=1$,
where $\mathbf{n}(\mathbb{\mathbf{k}})\equiv\frac{\mathbf{R}(\mathbf{k})}{|\mathbf{R}(\mathbf{k})|}$ and where the integrand divided by two is called the Berry curvature.

Now, we show how to obtain a scale-invariant cMERA for this model. 

\textit{Entanglement renormalization of the Chern insulator.\textemdash{}}Following the convention in Refs.\ \cite{Nozaki2012,Haegeman2013,Wen2016},
we take the Gaussian ansatz for the disentangler in the Schr\"odinger
picture, $K(u)=i\int \mathrm{d}^{2}\mathbf{k}\Big[g(\mathbf{k},u)\psi_{1}^{\dagger}(\mathbf{k})\psi_{2}(\mathbf{k})+g^{*}(\mathbf{k},u)\psi_{1}(\mathbf{k})\psi_{2}^{\dagger}(\mathbf{k})\Big]$ 
\footnote{In the cMERA literature,
a momentum cutoff $\Lambda$ is typically provided \cite{Haegeman2013,Wen2016}. With a finite cutoff, the UV state generated by a cMERA circuit approximates the ground state of the Hamiltonian up to $O(\frac{1}{\Lambda})$ corrections. Here, we work in the continuum limit $\Lambda \rightarrow \infty$ to avoid this technical subtlety. In principle, these finite-$\Lambda$ corrections can be worked out explicitly.}.
If we require our disentangler to be scale-invariant, then $g(\mathbf{k},u)$
should not have explicit $u$ dependence, $g(\mathbf{k},u)=g(\mathbf{k})$.
We also take the ansatz that $g(\mathbf{k})=\mathcal{H}(k)e^{-i\theta_{k}}$,
where $\mathcal{H}(k)$ is a real-valued function to be
determined. Through rewriting the disentangler as $K(u)=\int \mathrm{d}^{2}\mathbf{k}\,\psi^{\dagger}(\mathbf{k})[ \mathbf{H}(\mathbf{k})\cdot\boldsymbol{\sigma}]\psi(\mathbf{k})$ with $\mathbf{H}(\mathbf{k})=(\mathcal{H}(k)\sin \theta_k,-\mathcal{H}(k)\cos \theta_k,0)$, we can intuitively understand its action by imagining an effective magnetic field of strength $\mathcal{H}(k)$ in a clockwise direction about the origin applied to the pseudo-spin at each momentum point. In the interaction picture, the disentangler
becomes
\begin{align}
\hat{K}(u)=i\int \mathrm{d}^{2}\mathbf{k} & \biggl[\mathcal{H}(e^{-u}k)e^{-i\theta_{k}}\psi_{1}^{\dagger}(\mathbf{k})\psi_{2}(\mathbf{k})\nonumber \\
 & +\mathcal{H}(e^{-u}k)e^{i\theta_{k}}\psi_{1}(\mathbf{k})\psi_{2}^{\dagger}(\mathbf{k})\biggr].\label{eq:interactiondisentangler}
\end{align}

\begin{figure}
\begin{centering}
\includegraphics[width=0.8\columnwidth]{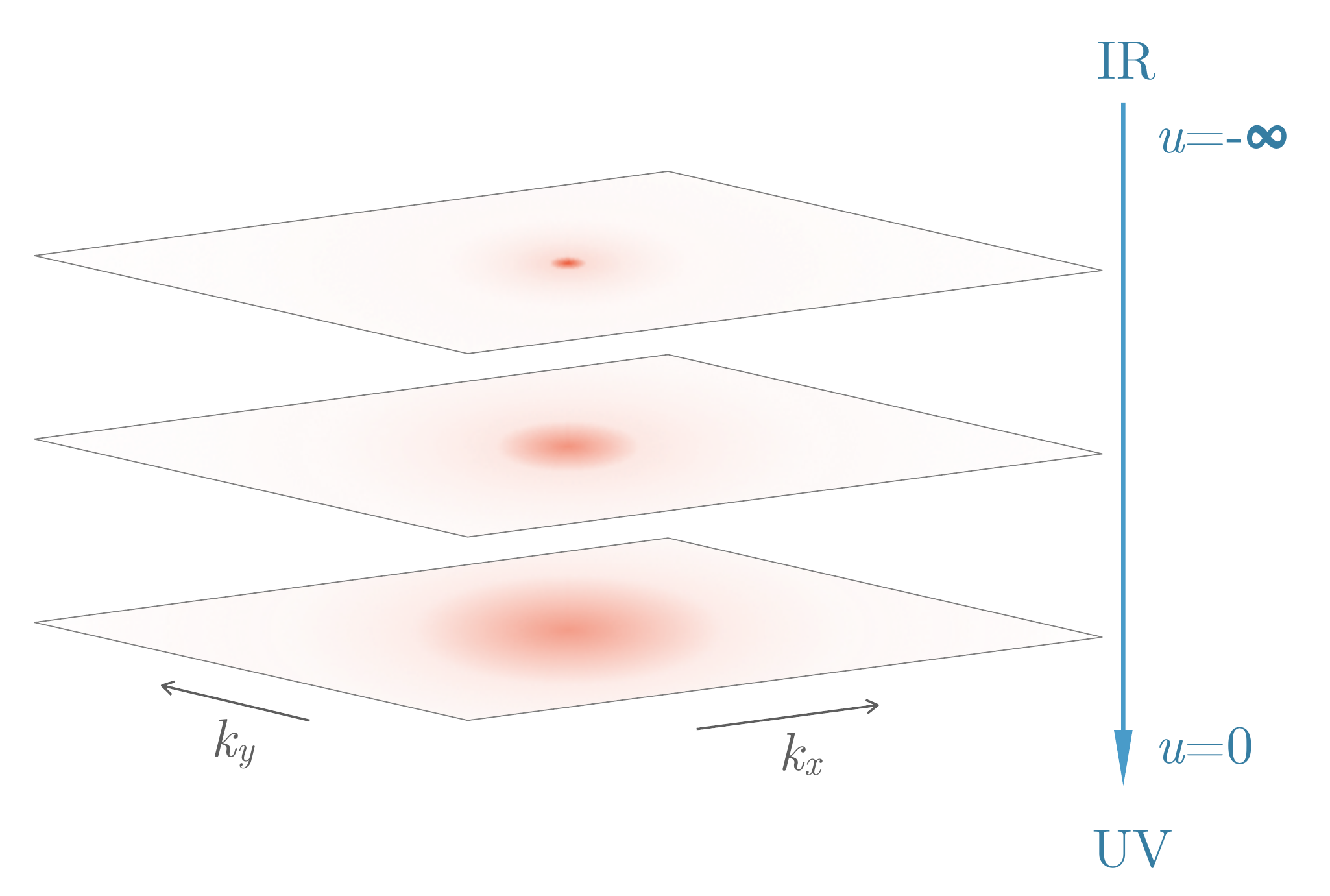}
\par\end{centering}
\centering{}\caption{Berry curvature of the renormalized wavefunction in the interaction
picture at different scales $u$, drawn schematically in momentum
space. The blue arrow corresponds to the direction of the reversed cMERA
circuit. The area contributing to the Chern number expands as one
approaches the UV scale.\label{fig:holography} }
\end{figure}

Now, we start to renormalize the wavefunction and determine the form
of the disentangler. We assume that the renormalized wavefunction at
scale $u$ can be expressed as
\begin{equation}
\ket{\Psi(u)}=\prod_{\mathbf{k}}(P_{\mathbf{k}}(u)\psi_{2}^{\dagger}(\mathbf{k})-Q_{\mathbf{k}}(u)\psi_{1}^{\dagger}(\mathbf{k}))\ket{\mathrm{vac}},\label{eq:renormalizedwavefunction}
\end{equation}
 with $\left|P_{\mathbf{k}}(u)\right|^{2}+\left|Q_{\mathbf{k}}(u)\right|^{2}=1$.
From Eq.\ (\ref{eq:interactioneq}), we get
\begin{align}
P_{\mathbf{k}}(u) & =A_{\mathbf{k}}e^{-i\varphi(e^{-u}\mathbf{k})}+B_{\mathbf{k}}e^{i\varphi(e^{-u}\mathbf{k})},\label{eq:PandQ}\\
Q_{\mathbf{k}}(u) & =-ie^{-i\theta_{k}}\left[A_{\mathbf{k}}e^{-i\varphi(e^{-u}\mathbf{k})}-B_{\mathbf{k}}e^{i\varphi(e^{-u}\mathbf{k})}\right].\nonumber 
\end{align}
Coefficients $A_{\mathbf{k}}$ and $B_{\mathbf{k}}$ are complex numbers
with $|A_{\mathbf{k}}|^{2}+|B_{\mathbf{k}}|^{2}=\frac{1}{2}$, and
$\varphi(e^{-u}\mathbf{k})\equiv\int_{ke^{-u}}^{\infty}\mathcal{H}\left(t\right)\frac{\mathrm{d}t}{t}$.
At UV scale $u=0$, the wavefunction should match the ground state
in Eq.\ (\ref{eq:groundstate}); at IR scale $u\to -\infty$, we would like the renormalized
wavefunction to be the product state $\prod_{\mathbf{k}}\psi_{1}^{\dagger}(\mathbf{k})\ket{\mathrm{vac}}$
or the product state $\prod_{\mathbf{k}}\psi_{2}^{\dagger}(\mathbf{k})\ket{\mathrm{vac}}$
\cite{Nozaki2012,Haegeman2013,Wen2016}. By taking $A_{\mathbf{k}}=-\frac{1}{2i}$
and $B_{\mathbf{k}}=\frac{1}{2i}$, the boundary conditions can be
satisfied by requiring
\begin{equation}
\mathcal{\mathcal{H}}\left(k\right)=\frac{k(m+k^{2})}{2\left[k^{4}+k^{2}(1-2m)+m^{2}\right]}.\label{eq:Hktheta}
\end{equation}
Substituting Eq.\ (\ref{eq:Hktheta}) into Eqs.\ (\ref{eq:renormalizedwavefunction})
and (\ref{eq:PandQ}), we attain an explicit form of the renormalized
wavefunction, 
\begin{align}
\ket{\Psi(u)}= & \prod_{\mathbf{k}}\frac{1}{\sqrt{N_{k,u}}}\times\nonumber \\
\bigg[\Big((m-k^{2}e^{-2u}) & +\sqrt{(m-k^{2}e^{-2u})^{2}+k^{2}e^{-2u}}\Big)\:\psi_{2}^{\dagger}(\mathbf{k})\nonumber \\
-k\,e^{-u}e^{-i\theta_{k}}\: & \psi_{1}^{\dagger}(\mathbf{k})\bigg]\ket{\mathrm{vac}},\label{eq:determinedrenormalizedwavefunction}
\end{align}
where $N_{k,u}$ is a normalization factor that depends on $k$ and
$u$. The Berry curvature of the renormalized wavefunction at different
$u$ is shown schematically in FIG. \ref{fig:holography}. The IR
state is $\ket{\Omega_{\mathrm{IR}}}=\lim_{u\rightarrow-\infty}\ket{\Psi(u)}=\prod_{\mathbf{k}}e^{-i\theta_{k}}\psi_{1}^{\dagger}(\mathbf{k})\ket{\mathrm{vac}}$,
which is equal to $\prod_{\mathbf{k}}\psi_{1}^{\dagger}(\mathbf{k})\ket{\mathrm{vac}}=\prod_{\mathbf{x}}\psi_{1}^{\dagger}(\mathbf{x})\ket{\mathrm{vac}}$
up to an overall phase. Note that the nonzero Chern number does not survive
in the IR state because the integration operation does not commute
with the limit $u\rightarrow-\infty$. However, at any finite $u$, the
Chern number is always one. Therefore, there is no phase transition
during the entanglement renormalization process, consistent with the
result in Ref.\ \cite{Wen2016}.

To analyze the spatial structure of the disentangler, we rewrite the expression
for $\mathcal{\mathcal{H}}\left(k\right)$. We first define
$\lambda_{+}$ and $\lambda_{-}$ as the two roots of the equation
$x^{2}+(1-2m)x+m^{2}=0$, $\lambda_{\pm}=\frac{-1+2m\pm\sqrt{1-4m}}{2}$.
They are real and negative when $0<m<1/4$. Although setting this
restriction on $m$ is not necessary for our disentangler, we will assume
it in the following in order to assist our experimental realization.
Now, the expression $\mathcal{\mathcal{H}}\left(k\right)$
can be rewritten as 
\begin{align}
\mathcal{\mathcal{H}}\left(k\right)= & \left(\frac{-1+\sqrt{1-4m}}{4\sqrt{1-4m}}\right)\frac{k}{k^{2}-\lambda_{+}}\nonumber \\
 & +\left(\frac{1+\sqrt{1-4m}}{4\sqrt{1-4m}}\right)\frac{k}{k^{2}-\lambda_{-}}.\label{eq:Hkthetadecom}
\end{align}
By inserting this expression into Eq.\ (\ref{eq:interactiondisentangler})
and performing a Fourier transform, it can be shown that the disentangler
in real space decays exponentially with characteristic length $e^{-u}\mathrm{max}\{\sqrt{-\lambda_{+}},\,\sqrt{-\lambda_{-}}\}$.
Therefore, our cMERA involves quasi-local
interactions.

\begin{figure}
\begin{centering}
\includegraphics[width=0.4\columnwidth]{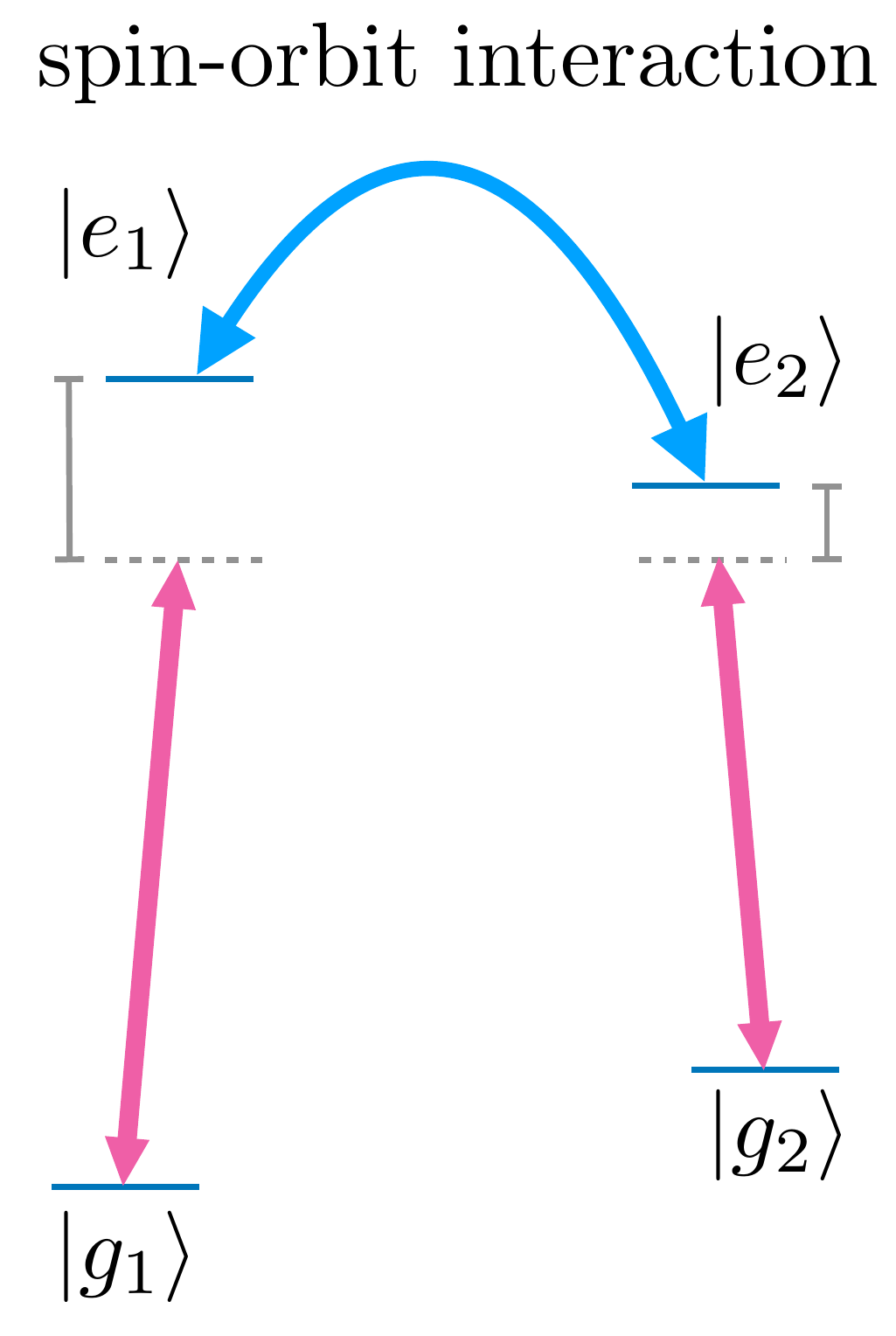}
\par\end{centering}
\centering{}\caption{A scheme to engineer the cMERA circuit in the interaction picture.
The two excited states are coupled by spin-orbit interaction to each other and by off-resonant lasers to the two ground states.  \label{fig:circuiteng} }
\end{figure}

\label{sec:realization}
\textit{Experimental realization of the cMERA circuit.\textemdash{}}We propose a  way to realize our reversed cMERA circuit to
prepare a Chern insulator state in an optical lattice with neutral $^{171}\mathrm{Yb}$,
which are fermionic atoms with two outer electrons.
From now on, we will drop the word ``reversed'' for our cMERA circuit
when the context is clear. Recall that the cMERA circuit starts with
an initial IR state. As discussed above, the IR state at $u\rightarrow-\infty$
does not have the correct Chern number; therefore, we
 start from a near-IR state with large negative $u$. In addition,
the cMERA circuit is only valid on a lattice when the continuum approximation 
holds. Therefore, throughout the circuit, the physical length scale $e^{-u} \mathrm{max}\{\sqrt{-\lambda_+},\sqrt{-\lambda_-}\}$ should be significantly larger than the lattice spacing, but significantly smaller than the total size of the lattice. Going forward, we begin
with a near-IR state and use our cMERA circuit to obtain the UV state
without ever violating the requirements of the continuum approximation.

Here, we assume that we already have an initial near-IR state waiting
to be inserted into the cMERA circuit. Since, in finite-size systems,
the Berry curvature is concentrated on a few discrete momentum points
near $k=0$, the preparation of this near-IR state should be fast
if we can individually create states at each point in momentum space.
In the Supplemental Material, we provide one possible method for generating
this initial state.

We now present the cMERA circuit engineering scheme (see Supplemental Material for details).
We use $\ket{g_{1}}$ and $\ket{g_{2}}$ as shorthand
notations for the two stable hyperfine ground states $\ket{F=1/2,\:m_{F}=-1/2}$
and $\ket{F=1/2,\:m_{F}=1/2}$ in $^{1}\mathrm{S}_{0}$; these
form the basis of our spinor $\psi(\mathbf{k})\equiv\left(\begin{array}{cc}
                                                            \psi_{1}(\mathbf{k}) & \psi_{2}(\mathbf{k})\end{array}\right)^{T}$.
We find that if we have two metastable excited states $\ket{e_1}$ and $\ket{e_2}$ (e.g.\ from the $^{3}$P  manifold) with quadratic dispersions coupled by spin-orbit
interaction and couple them off-resonantly to the respective ground
states as shown in FIG. \ref{fig:circuiteng}, then the disentangler
in the interaction picture can be engineered. Intuitively, the spin-orbit interaction allows us to generate a momentum-dependent effective magnetic field for Eq.\ (\ref{eq:interactiondisentangler}), whereas the off-resonant couplings to quadratic dispersive bands induce quadratic terms in the denominators of Eq.\ (\ref{eq:Hkthetadecom}). To accomplish this,
we utilize the scheme detailed in Refs.\ \cite{Campbell2011,Campbell2016,Huang2015a,Galitski2013}
to create two dressed excited states coupled by spin-orbit interaction.
However, as the two dressed states are linear combinations of bare
excited states, the dressed states do not have good quantum numbers
to have clear selection rules to forbid the transitions $\ket{g_{1}}\longleftrightarrow\ket{e_{2}}$
and $\ket{g_{2}}\longleftrightarrow\ket{e_{1}}$. Nevertheless, by carefully choosing the driving fields to couple ground states to the bare excited
states, we can create interferences to generate synthetic selection
rules. By varying the laser parameters as the circuit progresses,
we can engineer the disentangler in the interaction picture.

When the UV state is generated by the cMERA circuit, one can then
use the experimental techniques introduced in Refs.\ \cite{Hauke2014,Alba:2011aa,Flaschner:2016aa}
to measure the Chern number and the Berry curvature.

\textit{Discussion.\textemdash{}}In this work, we found a quasi-local cMERA whose fixed-point wavefunction
is a Chern insulator. This is a novel and unexpected
way to represent systems with chiral topological
order.
We also demonstrate that our quasi-local quantum circuit can 
be realized experimentally in a cold atom system, despite the common intuition
that a quantum circuit should be strictly local to allow easier implementation.

In our realization, we only explored one possibility to engineer spin-orbit
coupling, but it may be possible to engineer the interaction in other
ways, such as using magnetic fields on a chip \cite{Anderson2013} or
microwaves \cite{Grusdt2017}.
Other alkaline-earth atoms could also provide
promising experimental platforms. Although our experimental realization took
place in the interaction picture, one could in principle use the Schr\"odinger
picture for cMERA, where the lattice constant must continuously contract
\cite{Huckans2009,Al-Assam2010}.

It is also interesting that the Chern insulator ground state
is a fixed point of our cMERA with \emph{finite} correlation length.
This observation seems to contradict  the usual intuition that
the fixed point correlation length must be zero or infinity, as the
correlation length must decrease under rescaling of each strictly
local RG step in real space. However, since our cMERA involves continuous
time evolution and quasi-local interactions, it has potential
to restore the original correlation length after a finite time evolution.
The no-go theorems in Refs.\ \cite{Barthel2010,Dubail2015,Wahl2013,Li2017}
are similarly circumvented by a cMERA construction. Our work suggests that quasi-local RG transformations
are a more powerful framework than strictly local
RG transformations. It also might shed light on some of the key properties
of MERA-like formalisms for a wide range of chiral topological states.
In the future, we hope to extend the methods of this Letter to fractional
quantum Hall states.

\begin{acknowledgments} 
We are grateful to Bela Bauer, Yu-Ting Chen, Ze-Pei Cian, Ignacio Cirac, Glen Evenbly, Zhexuan Gong, Norbert Schuch, Brian Swingle, Tsz-Chun
Tsui, Brayden Ware, and Xueda Wen for helpful discussions. This project is supported
by the AFOSR, NSF QIS, ARL CDQI, ARO MURI, ARO, NSF PFC at JQI, and NSF Ideas Lab.
S.K.C. partially completed this work during his participation in the long-term workshop ``Entanglement in Quantum Systems'' held at the Galileo Galilei Institute for Theoretical Physics as well as ``Boulder School 2018: Quantum Information,'' which is supported by the National Science Foundation and the University of Colorado. He is also funded by the ACRI fellowship under the Young Investigator Training Program 2017. G.Z. is also supported by ARO-MURI, YIP-ONR and NSF CAREER (DMR431753240). J.R.G. acknowledges
support from the NIST NRC Research Postdoctoral Associateship Award.
Z.E. is supported in part by the ARCS Foundation. I.B.S. and A.V.C.
 acknowledge the additional support of the AFOSR's Quantum Matter MURI and
NIST. 
\end{acknowledgments}


%

\clearpage
\onecolumngrid 
\setcounter{figure}{0} 
\makeatletter 
\renewcommand{\thefigure}{S\@arabic\c@figure} 
\setcounter{equation}{0} 
\makeatletter
\renewcommand 
\theequation{S\@arabic\c@equation} 
\renewcommand
\thetable{S\@arabic\c@table}
\begin{center} 
{\large \bf Supplemental Material}
\end{center}

In this Supplemental Material, we provide details on the experimental realization.
In Section \ref{appsec:syntheticselectionrules}, we show how to engineer a synthetic selection rule between dressed states in the absence of any good quantum number. With that technique in mind, we
show a scheme to realize the cMERA circuit in Section \ref{appsec:summary}.
After that, in Section \ref{appsec:IR}, we provide one way to prepare
the initial state for the cMERA circuit by using spatial light modulators \cite{Fukuhara:2013aa,Nogrette2014}.

\section{Synthetic Selection Rules}

\label{appsec:syntheticselectionrules}

\begin{figure}
\centering{}\includegraphics[scale=0.3]{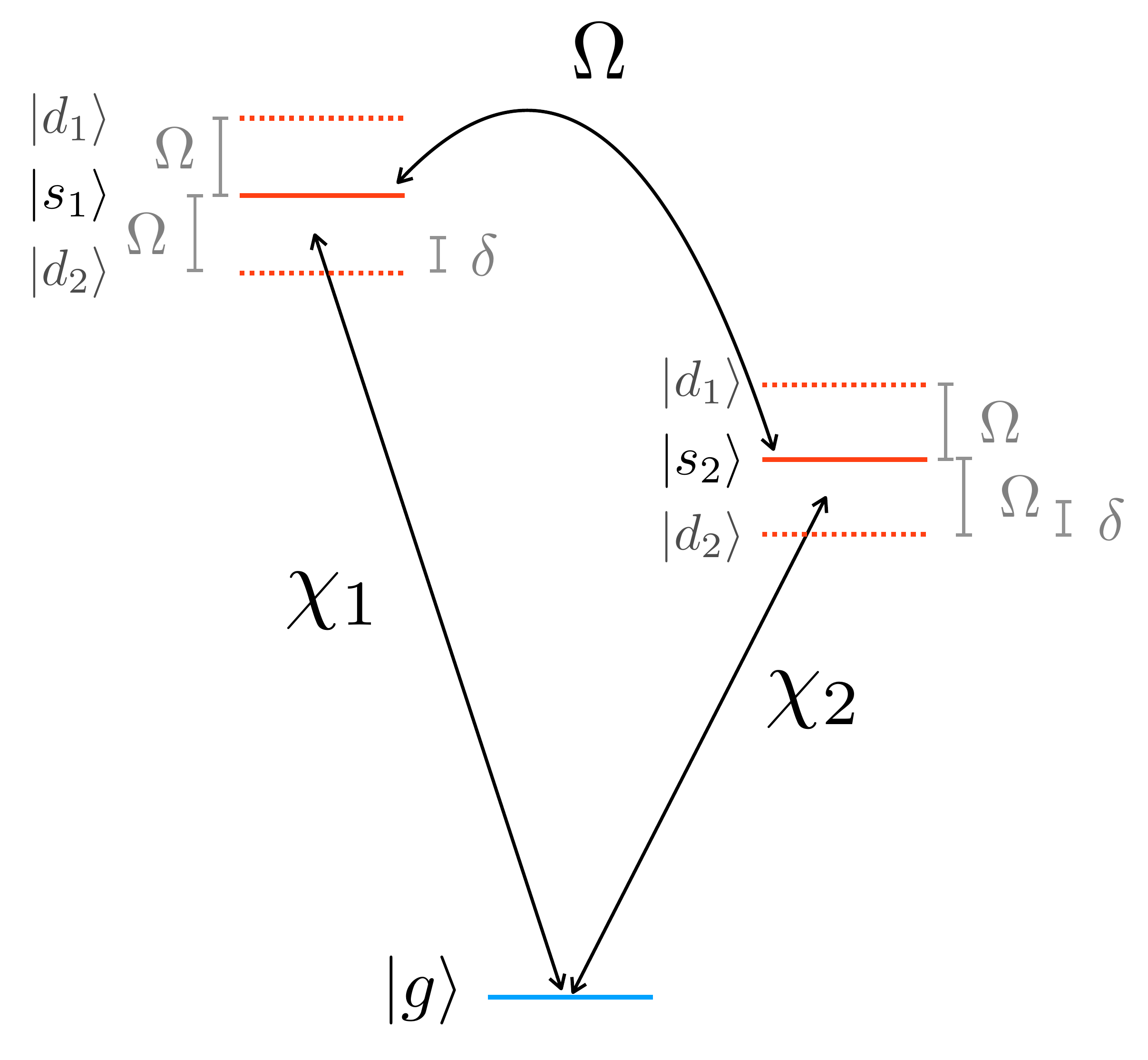}\caption{A toy model of synthetic selection rules. Bare states $\ket{s_{1}}$
and $\ket{s_{2}}$ are driven by a field with Rabi frequency $\Omega$,
whereby two dressed states $\ket{d_{1}}$ and $\ket{d_{2}}$ are created.
In view of the rotating frame, the dressed states are linear combinations
of bare states. As a result, they do not have good quantum numbers
to constitute a selection rule when coupling to another state, say
$\ket{g}$. A synthetic selection rule can be generated through applying
two driving fields from $\ket{g}$ to $\ket{s_{1}}$ and $\ket{s_{2}}$
with fine-tuned Rabi frequencies $\chi_{1}$ and $\chi_{2}$, respectively.
For example, we can forbid the transition from $\ket{g}$ to $\ket{d_{1}}$
by choosing $\chi_{1}=-\chi_{2}$. \label{appfig:syntheticselectionrules} }
\end{figure}

In this section, we introduce a trick that will be useful
for engineering the disentangler in a real atomic system. Suppose
that we have a three-level system composed of states $\ket{s_{1}}$,
$\ket{s_{2}}$, and $\ket{g}$. In the presence of an on-resonance
driving with Rabi frequency $\Omega$ between bare states $\ket{s_{1}}$
and $\ket{s_{2}}$, two dressed states $\ket{d_{1}}$ and $\ket{d_{2}}$ are formed.
We are going to show that by fine-tuning
the Rabi frequencies $\chi_{1}$ and $\chi_{2}$, we can
generate a synthetic selection rule from state $\ket{g}$ to the two
dressed states $\ket{d_{1}}$ and $\ket{d_{2}}$, e.g., $\ket{g}\rightarrow\ket{d_{2}}$
is allowed while $\ket{g}\rightarrow\ket{d_{1}}$ is forbidden. (Once
we prove this, the converse case where $\ket{g}\rightarrow\ket{d_{1}}$
is allowed and $\ket{g}\rightarrow\ket{d_{2}}$ is forbidden is a trivial generalization.) We
consider a driving Hamiltonian, which under rotating wave approximation
is
\[
h=\left(\begin{array}{ccc}
0 & \chi_{1}^{*}e^{i(\omega_{1}-\Omega+\delta)t} & \chi_{2}^{*}e^{i(\omega_{2}-\Omega+\delta)t}\\
\chi_{1}e^{-i(\omega_{1}-\Omega+\delta)t} & \omega_{1} & \Omega e^{-i(\omega_{1}-\omega_{2})t}\\
\chi_{2}e^{-i(\omega_{2}-\Omega+\delta)t} & \Omega e^{i(\omega_{1}-\omega_{2})t} & \omega_{2}
\end{array}\right).
\]
The order of the columns (rows) is $\ket{g}$, $\ket{s_{1}}$, $\ket{s_{2}}$.
We have assumed that $|\omega_{1}-\omega_{2}|\gg\Omega$, allowing us to neglect some transitions that are far off-resonant. The level
diagram is illustrated in FIG. \ref{appfig:syntheticselectionrules}.

Going to the rotating frame defined by the unitary matrix
\[
U=\left(\begin{array}{ccc}
1 & 0 & 0\\
0 & e^{-i(\omega_{1}-\omega_{2})t} & 0\\
0 & 0 & 1
\end{array}\right),
\]
we obtain the effective Hamiltonian
\[
U^{\dagger}hU-i\partial_{t}U^{\dagger}U=\left(\begin{array}{ccc}
0 & \chi_{1}^{*}e^{i(\omega_{2}-\Omega+\delta)t} & \chi_{2}^{*}e^{i(\omega_{2}-\Omega+\delta)t}\\
\chi_{1}e^{-i(\omega_{2}-\Omega+\delta)t} & \omega_{2} & \Omega\\
\chi_{2}e^{-i(\omega_{2}-\Omega+\delta)t} & \Omega & \omega_{2}
\end{array}\right).
\]
After diagonalizing the $2\times2$ block on the bottom right, we obtain the following Hamiltonian:
\[
\left(\begin{array}{ccc}
0 & \frac{1}{\sqrt{2}}(\chi_{1}^{*}+\chi_{2}^{*})e^{i(\omega_{2}-\Omega+\delta)t} & \frac{1}{\sqrt{2}}(\chi_{1}^{*}-\chi_{2}^{*})e^{i(\omega_{2}-\Omega+\delta)t}\\
\frac{1}{\sqrt{2}}(\chi_{1}+\chi_{2})e^{-i(\omega_{2}-\Omega+\delta)t} & \omega_{2}+\Omega & 0\\
\frac{1}{\sqrt{2}}(\chi_{1}-\chi_{2})e^{-i(\omega_{2}-\Omega+\delta)t} & 0 & \omega_{2}-\Omega
\end{array}\right).
\]

We denote the dressed state with energy $\omega_{2}+\Omega$ as $\ket{d_{1}}$,
and the dressed state with energy $\omega_{2}-\Omega$ as $\ket{d_{2}}$.
We can see that if  we fine-tune $\chi_{1}=-\chi_{2}$, we synthesize a selection rule where only the transition
between $\ket{d_{2}}$ and $\ket{g}$ is allowed. The synthetic Rabi frequency is then $\sqrt{2}\chi_{1}$.

This synthetic selection rule can be understood by considering two separate rotating frames with respect to states $\ket{s_{1}}$ and $\ket{s_{2}}$, as shown in FIG. \ref{appfig:syntheticselectionrules}.
In each rotating frame, we have dressed states $\ket{d_{1}}$ and
$\ket{d_{2}}$. We can couple $\ket{g}$ to dressed states either
by driving $\ket{g}$ to dressed states in the $\ket{s_{1}}$ rotating
frame or in the $\ket{s_{2}}$ rotating frame. By creating interference 
between the two channels, we obtain a synthetic selection rule.

\section{The Continuous MERA Circuit Engineering }

\label{appsec:summary}

\begin{figure}
\begin{centering}
(a)\includegraphics[width=0.6\columnwidth]{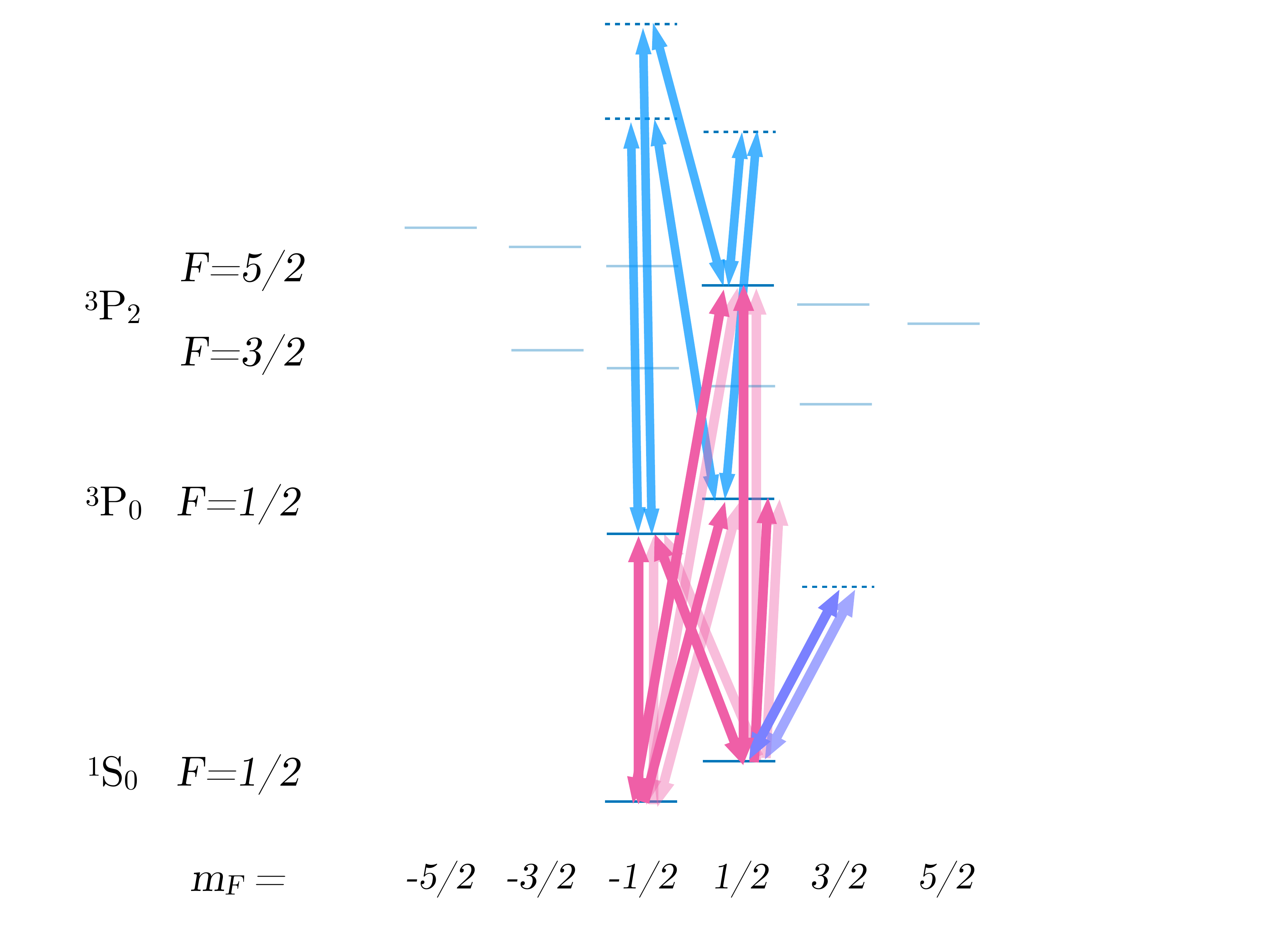}(b)\includegraphics[width=0.3\columnwidth]{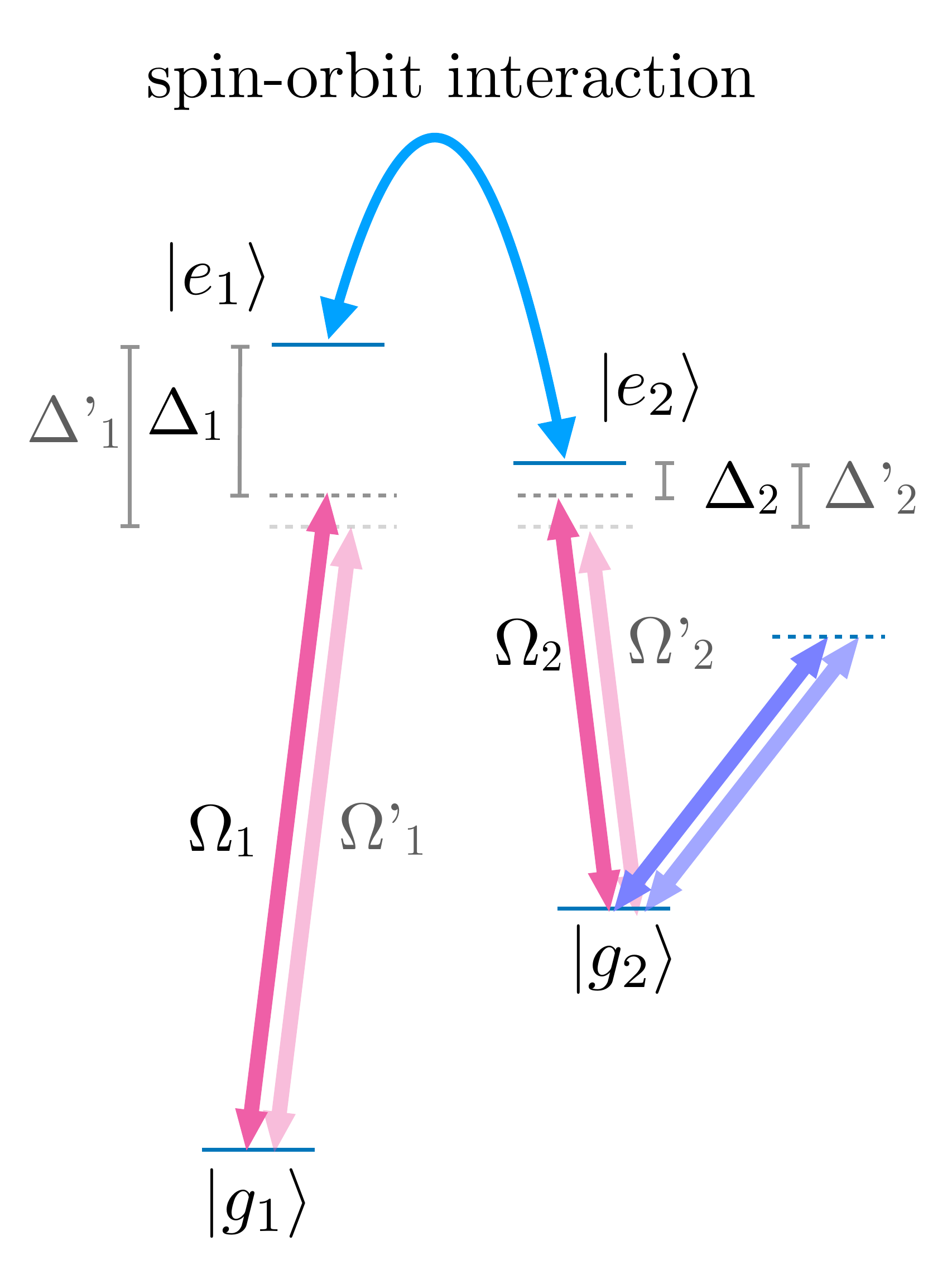}
\par\end{centering}
\centering{}\caption{Disentangler engineering. (a) A magnetic field is applied to induce
hyperfine splittings. The excited states are coupled by Raman beams (colored in blue)
to generate an effective spin-orbit interaction. They are chosen from the hyperfine manifolds  $^{3}\mathrm{P}_2$ $F=5/2$ and   $^{3}\mathrm{P}_0$ $F=1/2$, which are long-lived to circumvent dissipation issues. Their ultra-narrow linewidths are on the order of tens of millihertz \cite{Zhang2015,Kohno2009,Lemke2009,Park2013,Yamaguchi2008}. Additionally, we also have two sets
of multiple lasers, colored in light and dark pink, coupling the ground
states to the excited states to engineer the disentangler of our cMERA
by creating synthetic selection rules. (b) The effective couplings
between ground states and the dressed excited states are generated
from the scheme shown in (a). We ignore a third dressed state since
it is far off-resonant. Now we effectively create two dressed
excited states coupled by spin-orbit interaction, which are coupled
to the ground states by two pairs of drivings colored in light and
dark pink. The synthetic selection rules forbid $\ket{g_{1},\mathbf{k}}\protect\longleftrightarrow\ket{e_{2},\mathbf{k}}$
and $\ket{g_{2},\mathbf{k}}\protect\longleftrightarrow\ket{e_{1},\mathbf{k}}$.
The effective Rabi frequencies and detunings for two pairs of effective
drivings are labeled by unprimed and primed notation. The band structures
are ignored in this picture, so by detunings we mean the detunings
at $\mathbf{k}=0$. The light and dark purple arrows on the bottom right in (a) and (b) both represent
lasers used to cancel unwanted AC Stark shifts by coupling the ground
states to some negative curvature bands of some excited state, e.g., an unused excited state in the $^3 \mathrm{P}_2$ $F=5/2$ hyperfine manifold.\label{appfig:engineer} }
\end{figure}
\begin{figure}
\begin{centering}
\includegraphics[width=0.5\columnwidth]{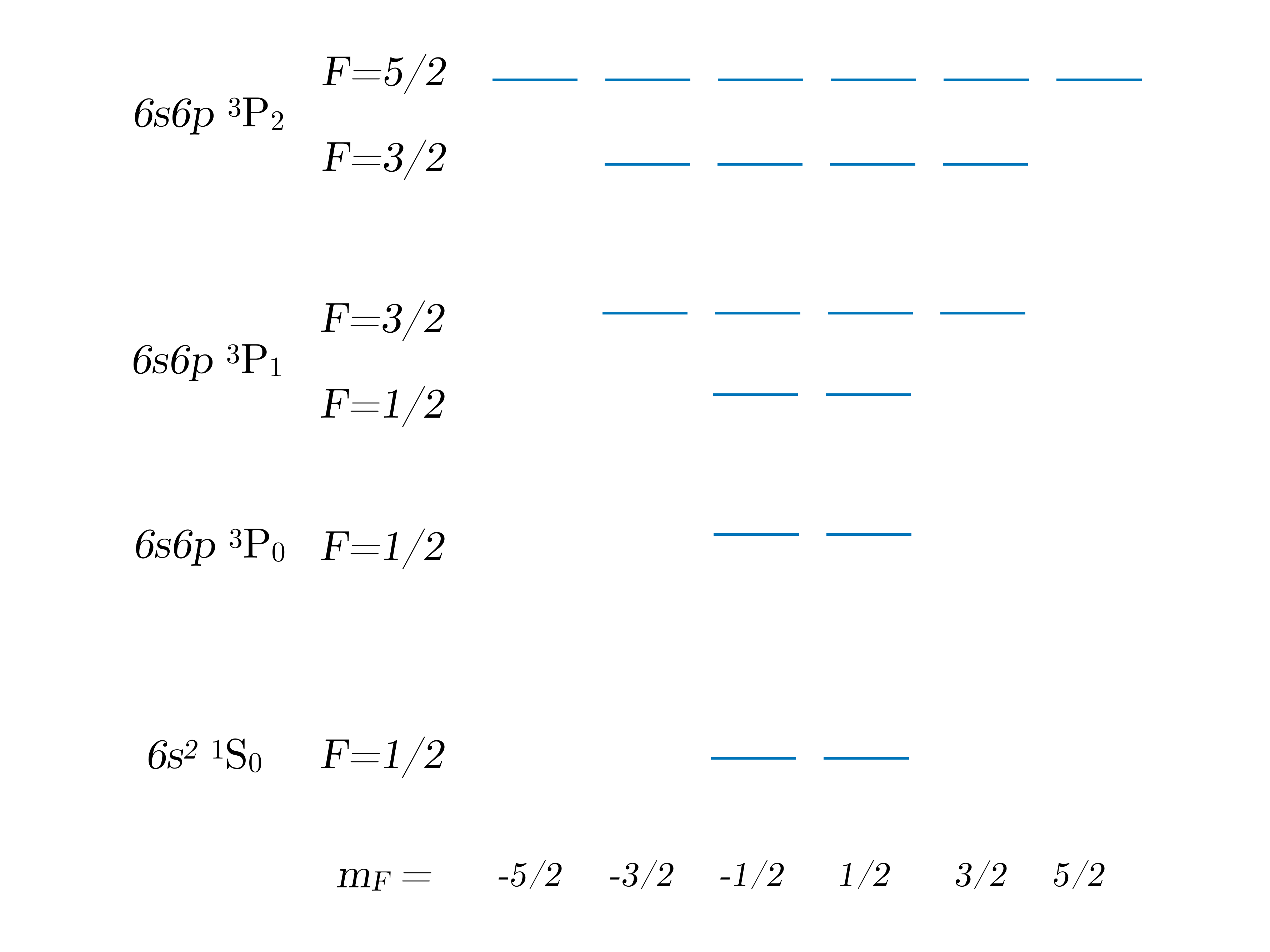}
\par\end{centering}
\centering{}\caption{Energy level diagram of neutral atom $^{171}\mathrm{Yb}$. The hyperfine
structure is shown. We employ the bottom two ground states as our
spinor basis of the Chern insulator. \label{appfig:level} }
\end{figure}

In this section, we show that by using the scheme
shown in FIG. \ref{appfig:engineer}(b), we can engineer the
disentangler in the interaction picture. Here, we choose the two hyperfine
ground states of $^{171}\mathrm{Yb}$ shown in FIG. \ref{appfig:level}
as our spinor basis of the Chern insulator and effectively couple
them to some dressed excited states by two pairs of driving fields. The
meaning of ``dressed'' excited states will become clear shortly.
Additionally, the dressed excited states are coupled by spin-orbit
interaction, while transitions $\ket{g_{1},\mathbf{k}}\longleftrightarrow\ket{e_{2},\mathbf{k}}$
and $\ket{g_{2},\mathbf{k}}\longleftrightarrow\ket{e_{1},\mathbf{k}}$
are forbidden. In order to implement this idea in neutral $^{171}\mathrm{Yb}$
atoms, we need to use techniques introduced in Refs.\ \cite{Campbell2011,Campbell2016}
and Section \ref{appsec:syntheticselectionrules}. To create states
coupled by spin-orbit coupling, we will utilize the method discussed
in Refs.\ \cite{Campbell2011,Campbell2016}. However, the dressed states created by that scheme do not
have good quantum numbers to enforce selection rules. Therefore, we use the technique outlined
in Section \ref{appsec:syntheticselectionrules} to create a synthetic selection rule. In this part of
the Supplemental Material, we show how to combine those
techniques consistently in neutral  $^{171}\mathrm{Yb}$.

First, we show how FIG. \ref{appfig:engineer}(b)
arises from FIG. \ref{appfig:engineer}(a), inducing the disentangler
interaction. We first consider the case with the set of lasers colored
in dark pink in FIG. \ref{appfig:engineer}(a) with additional Raman lasers
coupling the bare excited states. This will give rise to the effective
unprimed pair of drivings in FIG. \ref{appfig:engineer}(b). We will
find that this scheme generates one term in our disentangler with
$\mathcal{\mathcal{H}}\left(k\right)$ described by Eq.\
(\ref{eq:Hkthetadecom}) in the main text. Therefore, to produce another
term, we will use another set of lasers with different parameters,
which will effectively induce the primed pair of drivings in FIG.
\ref{appfig:engineer}(b).

We assume that states $\ket{g_{1}}$ and $\ket{g_{2}}$ have flat
bands, whereas the chosen bare excited states are weakly trapped.
In the continuum, low-energy limit, atoms in the bare excited states
can be described by non-relativistic particles with mass $M$. After
appropriate Raman transitions for the bare excited states, we obtain
the effective Hamiltonian in the rotating frame of the basis $\ket{g_{1}}$, $\ket{g_{2}}$, $\ket{e_{\mathrm{bare},1}}$,
$\ket{e_{\mathrm{bare},2}}$, and $\ket{e_{\mathrm{bare},3}}$
under the rotating wave approximation:
\[
\begin{array}{c}
h=\left(\begin{array}{ccccc}
0 & 0 & \chi_{1,1}^{*}e^{i\Delta t} & \chi_{1,2}^{*}e^{i\Delta t} & \chi_{1,3}^{*}e^{i\Delta t}\\
0 & 0 & \chi_{2,1}^{*}e^{i\Delta t} & \chi_{2,2}^{*}e^{i\Delta t} & \chi_{2,3}^{*}e^{i\Delta t}\\
\chi_{1,1}e^{-i\Delta t} & \chi_{2,1}e^{-i\Delta t} & \frac{(\mathbf{k}+\mathbf{k}_{1})^{2}}{2M} & \Omega e^{i\phi_{1,2}} & \Omega e^{-i\phi_{3,1}}\\
\chi_{1,2}e^{-i\Delta t} & \chi_{2,2}e^{-i\Delta t} & \Omega e^{-i\phi_{1,2}} & \frac{(\mathbf{k}+\mathbf{k}_{2})^{2}}{2M} & \Omega e^{i\phi_{2,3}}\\
\chi_{1,3}e^{-i\Delta t} & \chi_{2,3}e^{-i\Delta t} & \Omega e^{i\phi_{3,1}} & \Omega e^{-i\phi_{2,3}} & \frac{(\mathbf{k}+\mathbf{k}_{3})^{2}}{2M}
\end{array}\right)\end{array}.
\]
The order of the columns is $\ket{g_{1},\mathbf{k}}$, $\ket{g_{2},\mathbf{k}}$
$\ket{e_{\mathrm{bare},1},\mathbf{k}+\mathbf{k}_{1}}$, $\ket{e_{\mathrm{bare},2},\mathbf{k}+\mathbf{k}_{2}}$,
and $\ket{e_{\mathrm{bare},3},\mathbf{k}+\mathbf{k}_{3}}$. The notation $\Delta$
is the common detuning of all the lasers coupling the two ground states to
the excited states, whereas $\chi_{i,j}$ represents the Rabi frequencies
of those lasers. We define the detuning at the zero momentum energy of the bare excited state. Here, $\mathbf{k}_{1}$,
$\mathbf{k}_{2}$, and $\mathbf{k}_{3}$ are subject to the condition
$|k_{1}|=|k_{2}|=|k_{3}|=k_{\mathrm{SOC}}$, $\mathbf{k}_{1}+\mathbf{k}_{2}+\mathbf{k}_{3}=0$,
and $\mathbf{k}_{j}=k_{\mathrm{SOC}}[\cos(2\pi j/3)\mathbf{e}_{x}+\sin(2\pi j/3)\mathbf{e}_{y}]$.

We apply the following unitaries to conjugate the single body Hamiltonian
\[
U=\left(\begin{array}{ccccc}
1 & 0 & 0 & 0 & 0\\
0 & 1 & 0 & 0 & 0\\
0 & 0 & e^{-i2\pi/3}/\sqrt{3} & e^{-i4\pi/3}/\sqrt{3} & 1/\sqrt{3}\\
0 & 0 & e^{-i4\pi/3}/\sqrt{3} & e^{-i8\pi/3}/\sqrt{3} & 1/\sqrt{3}\\
0 & 0 & s1/\sqrt{3} & 1/\sqrt{3} & 1/\sqrt{3}
\end{array}\right),
\]
\[
U'=\left(\begin{array}{ccccc}
1 & 0 & 0 & 0 & 0\\
0 & 1 & 0 & 0 & 0\\
0 & 0 & e^{i\left(\phi_{1,2}+\phi_{2,3}+\phi_{3,1}\right)/3} & 0 & 0\\
0 & 0 & 0 & e^{i\left(-\phi_{1,2}+2\phi_{2,3}+2\phi_{3,1}\right)/3} & 0\\
0 & 0 & 0 & 0 & e^{i\phi_{3,1}}
\end{array}\right),
\]
and assume the following to obtain a synthetic selection rule:
\begin{align}
\chi_{1,2} & =e^{2\pi i/3}e^{-i\,(2\phi_{1,2}-\phi_{2,3}-\phi_{3,1})/3}\chi_{1,1}\nonumber \\
\chi_{1,3} & =e^{-2\pi i/3}e^{-i\,(\phi_{1,2}+\phi_{2,3}-2\phi_{3,1})/3}\chi_{1,1}\nonumber \\
\chi_{2,1} & =e^{2\pi i/3}e^{i\,(2\phi_{1,2}-\phi_{2,3}-\phi_{3,1})/3}\chi_{2,2}\nonumber \\
\chi_{2,3} & =e^{-2\pi i/3}e^{i\,(\phi_{1,2}-2\phi_{2,3}+\phi_{3,1})/3}\chi_{2,2}.
\end{align}
The Hamiltonian becomes
\begin{equation}
(U'U)^{\dagger}hU'U=\left(\begin{array}{ccccc}
0 & 0 & \Omega_{1}^{*}e^{i\Delta t} & 0 & 0\\
0 & 0 & 0 & \Omega_{2}^{*}e^{i\Delta t} & 0\\
\Omega_{1}e^{-i\Delta t} & 0 & \frac{k^{2}+k_{\mathrm{SOC}}^{2}}{2M}+2\Omega\cos(\frac{2\pi}{3}-\phi) & \frac{k_{\mathrm{SOC}}}{M}(k_{x}-ik_{y}) & \frac{k_{\mathrm{SOC}}}{M}(k_{x}+ik_{y})\\
0 & \Omega_{2}e^{-i\Delta t} & \frac{k_{\mathrm{SOC}}}{M}(k_{x}+ik_{y}) & \frac{k^{2}+k_{\mathrm{SOC}}^{2}}{2M}+2\Omega\cos(\frac{4\pi}{3}-\phi) & \frac{k_{\mathrm{SOC}}}{M}(k_{x}-ik_{y})\\
0 & 0 & \frac{k_{\mathrm{SOC}}}{M}(k_{x}-ik_{y}) & \frac{k_{\mathrm{SOC}}}{M}(k_{x}+ik_{y}) & \frac{k^{2}+k_{\mathrm{SOC}}^{2}}{2M}+2\Omega\cos(\phi)
\end{array}\right),
\end{equation}
where $\Omega_{1}\equiv-\sqrt{3}e^{-i\pi/3}e^{-i(\phi_{1,2}+\phi_{2,3}+\phi_{3,1})/3}\chi_{1,1}$,
$\Omega_{2}\equiv-\sqrt{3}e^{-i\pi/3}e^{i(\phi_{1,2}-2\phi_{2,3}-2\phi_{3,1})/3}\chi_{2,2}$,
and $\phi\equiv(\phi_{1,2}+\phi_{2,3}+\phi_{3,1})/3$. The order of
the columns is $\ket{g_{1},\mathbf{k}}$, $\ket{g_{2},\mathbf{k}}$,
$\ket{e_{1},\mathbf{k}}$, $\ket{e_{2},\mathbf{k}}$, and $\ket{e_{3},\mathbf{k}}$.
States $\ket{e_{1},\mathbf{k}}$, $\ket{e_{2},\mathbf{k}}$, $\ket{e_{3},\mathbf{k}}$
are dressed excited states which are linear combinations of the bare
excited states $\ket{e_{\mathrm{bare},1},\mathbf{k}+\mathbf{k}_{1}}$, $\ket{e_{\mathrm{bare},2},\mathbf{k}+\mathbf{k}_{2}}$,
and $\ket{e_{\mathrm{bare},3},\mathbf{k}+\mathbf{k}_{3}}$. By adiabatically
eliminating the dressed excited state representing the third column
(row) to the zeroth order and expanding $\phi$ to the first order,
we obtain the effective Hamiltonian
\begin{equation}
\left(\begin{array}{cccc}
0 & 0 & \Omega_{1}^{*}e^{i\Delta t} & 0\\
0 & 0 & 0 & \Omega_{2}^{*}e^{i\Delta t}\\
\Omega_{1}e^{-i\Delta t} & 0 & \frac{k^{2}}{2M}+E_{\mathrm{SOC}}+\sqrt{3}\Omega\phi & \frac{k_{\mathrm{SOC}}}{M}(k_{x}-ik_{y})\\
0 & \Omega_{2}e^{-i\Delta t} & \frac{k_{\mathrm{SOC}}}{M}(k_{x}+ik_{y}) & \frac{k^{2}}{2M}+E_{\mathrm{SOC}}-\sqrt{3}\Omega\phi
\end{array}\right),
\end{equation}
where $E_{\mathrm{SOC}}\equiv k_{\mathrm{SOC}}^{2}/2M-\Omega$. The
order of the columns is $\ket{g_{1},\mathbf{k}}$, $\ket{g_{2},\mathbf{k}}$,
$\ket{e_{1},\mathbf{k}}$, and $\ket{e_{2},\mathbf{k}}$.
By inspecting the matrix elements, one can see that a spin-orbit
interaction and a synthetic selection rule shown in FIG. \ref{appfig:engineer}(b)
have been consistently generated as we claimed.

Now, we are going to show that with this Hamiltonian, we can almost
generate the disentangler.  First, we go to a frame in which $\ket{e_1,\mathbf{k}}$ and $\ket{e_2,\mathbf{k}}$ rotate with frequency $\Delta$. The Hamiltonian becomes
\[
\left(\begin{array}{cccc}
0 & 0 & \Omega_{1}^{*} & 0\\
0 & 0 & 0 & \Omega_{2}^{*}\\
\Omega_{1} & 0 & \frac{k^{2}}{2M}+E_{\mathrm{SOC}}-\Delta+\sqrt{3}\Omega\phi & \frac{k_{\mathrm{SOC}}}{M}(k_{x}-ik_{y})\\
0 & \Omega_{2} & \frac{k_{\mathrm{SOC}}}{M}(k_{x}+ik_{y}) & \frac{k^{2}}{2M}+E_{\mathrm{SOC}}-\Delta-\sqrt{3}\Omega\phi
\end{array}\right).
\]
For the sake of later convenience, we denote $\Delta_{1}\equiv E_{\mathrm{SOC}}-\Delta+\sqrt{3}\Omega\phi$
and $\Delta_{2}\equiv E_{\mathrm{SOC}}-\Delta-\sqrt{3}\Omega\phi$:
\begin{equation}
\left(\begin{array}{cccc}
0 & 0 & \Omega_{1}^{*} & 0\\
0 & 0 & 0 & \Omega_{2}^{*}\\
\Omega_{1} & 0 & \Delta_{1}+k^{2}/2M & \frac{k_{\mathrm{SOC}}}{M}(k_{x}-ik_{y})\\
0 & \Omega_{2} & \frac{k_{\mathrm{SOC}}}{M}(k_{x}+ik_{y}) & \Delta_{2}+k^{2}/2M
\end{array}\right).
\end{equation}
We can see that $\Delta_{1}$ and $\Delta_{2}$ correspond to the
effective detunings at $\mathbf{k}=0$. Define $\alpha=k_{\mathrm{SOC}}/M$
and $k$, $\theta_{k}$ such that $k\cos\theta_{k}=k_{x}$ and $k\sin\theta_{k}=k_{y}$
to simplify our calculations. Notice that we have chosen a different
sign convention of the detunings $\Delta_{1}$ and $\Delta_{2}$ from
the normal convention. We will assume that $\Delta_{1},\Delta_{2}>0$
in our system so that the effective drivings are red-detuned. Now
we conjugate the Hamiltonian with the following unitary matrix:
\begin{equation}
\left(\begin{array}{cccc}
1 & 0 & 0 & 0\\
0 & 1 & 0 & 0\\
0 & 0 & 1-\frac{\alpha^{2}k^{2}}{2(\Delta_{1}-\Delta_{2})^{2}} & -\frac{\alpha ke^{-i\theta_{k}}}{\Delta_{1}-\Delta_{2}}\\
0 & 0 & \frac{\alpha ke^{i\theta_{k}}}{\Delta_{1}-\Delta_{2}} & 1-\frac{\alpha^{2}k^{2}}{2(\Delta_{1}-\Delta_{2})^{2}}
\end{array}\right)+O\left(\left(\frac{\alpha k}{\Delta_{1}-\Delta_{2}}\right)^{3}\right),
\end{equation}
and the effective Hamiltonian to order $\left(\frac{\alpha k}{\Delta_{1}-\Delta_{2}}\right)^{3}$
becomes
\begin{equation}
\left(\begin{array}{cccc}
0 & 0 & \Omega_{1}^{*}\left(1-\frac{\alpha^{2}k^{2}}{2(\Delta_{1}-\Delta_{2})^{2}}\right) & -\frac{\Omega_{1}^{*}\alpha ke^{-i\theta_{k}}}{\Delta_{1}-\Delta_{2}}\\
0 & 0 & \frac{\Omega_{2}^{*}\alpha ke^{i\theta_{k}}}{\Delta_{1}-\Delta_{2}} & \Omega_{2}^{*}\left(1-\frac{\alpha^{2}k^{2}}{2(\Delta_{1}-\Delta_{2})^{2}}\right)\\
\Omega_{1}\left(1-\frac{\alpha^{2}k^{2}}{2(\Delta_{1}-\Delta_{2})^{2}}\right) & \frac{\Omega_{2}\alpha ke^{-i\theta_{k}}}{\Delta_{1}-\Delta_{2}} & \Delta_{1}+\frac{\alpha^{2}k^{2}}{\Delta_{1}-\Delta_{2}}+k^{2}/2M & 0\\
-\frac{\Omega_{1}\alpha ke^{i\theta_{k}}}{\Delta_{1}-\Delta_{2}} & \Omega_{2}\left(1-\frac{\alpha^{2}k^{2}}{2(\Delta_{1}-\Delta_{2})^{2}}\right) & 0 & \Delta_{2}-\frac{\alpha^{2}k^{2}}{\Delta_{1}-\Delta_{2}}+k^{2}/2M
\end{array}\right).
\end{equation}
If we assume that $M\gg\frac{k_{\mathrm{SOC}}^{2}}{\Delta_{1}-\Delta_{2}}$, we can ignore the terms $\frac{\alpha^{2}k^{2}}{\Delta_{1}-\Delta_{2}}$
in the $(3,3)$ and $(4,4)$ entries.
Now, we also drop $O\left(\left(\frac{\alpha k}{\Delta_{1}-\Delta_{2}}\right)^{2}\right)$ terms in the $(1,3)$, $(2,4)$, $(3,1)$, and $(4,2)$ entries.
The remaining Hamiltonian is 
\begin{equation}
\left(\begin{array}{cccc}
0 & 0 & \Omega_{1}^{*} & -\frac{\Omega_{1}^{*}\alpha ke^{-i\theta_{k}}}{\Delta_{1}-\Delta_{2}}\\
0 & 0 & \frac{\Omega_{2}^{*}\alpha ke^{i\theta_{k}}}{\Delta_{1}-\Delta_{2}} & \Omega_{2}^{*}\\
\Omega_{1} & \frac{\Omega_{2}\alpha ke^{-i\theta_{k}}}{\Delta_{1}-\Delta_{2}} & \Delta_{1}+k^{2}/2M & 0\\
-\frac{\Omega_{1}\alpha ke^{i\theta_{k}}}{\Delta_{1}-\Delta_{2}} & \Omega_{2} & 0 & \Delta_{2}+k^{2}/2M
\end{array}\right).
\end{equation}
We adiabatically eliminate the state in the first and second columns
(rows). The remaining Hamiltonian of the subspace spanned by dressed
states $\ket{\tilde{g_{1}},\mathbf{k}}$, and $\ket{\tilde{g_{2}},\mathbf{k}}$
is
\begin{equation}
\label{appeq:disengbeforemanyapprox}
\left(\begin{array}{cc}
-\frac{\left|\Omega_{1}\right|^{2}}{\Delta_{1}+k^{2}/2M}-\frac{\left|\Omega_{1}\right|^{2}}{\Delta_{2}+k^{2}/2M}\left(\frac{\alpha k}{\Delta_{1}-\Delta_{2}}\right)^{2} & \frac{\alpha ke^{-i\theta_{k}}\text{\ensuremath{\Omega_{1}^{*}}}\text{\ensuremath{\Omega_{2}}}}{(\Delta_{1}-\Delta_{2})(\Delta_{1}+k^{2}/2M)}-\frac{\alpha ke^{-i\theta_{k}}\text{\ensuremath{\Omega_{1}^{*}}}\text{\ensuremath{\Omega_{2}}}}{(\Delta_{1}-\Delta_{2})(\Delta_{2}+k^{2}/2M)}\\
\frac{\alpha ke^{i\theta_{k}}\text{\ensuremath{\Omega_{1}}}\text{\ensuremath{\Omega_{2}^{*}}}}{(\Delta_{1}-\Delta_{2})(\Delta_{1}+k^{2}/2M)}-\frac{\alpha ke^{i\theta_{k}}\text{\ensuremath{\Omega_{1}}}\text{\ensuremath{\Omega_{2}^{*}}}}{(\Delta_{1}-\Delta_{2})(\Delta_{2}+k^{2}/2M)} & -\frac{\left|\Omega_{2}\right|^{2}}{\Delta_{1}+k^{2}/2M}\left(\frac{\alpha k}{\Delta_{1}-\Delta_{2}}\right)^{2}-\frac{\left|\Omega_{2}\right|^{2}}{\Delta_{2}+k^{2}/2M}
\end{array}\right).
\end{equation}
We have assumed $\Delta_{1},\,\Delta_{2},\gg\Omega_{1},\:\Omega_{2}$.
A necessary condition of this assumption is that $\Omega\gg\Omega_{1},\:\Omega_{2}$.
Now, supposing that we can tune $\Delta_{1}\gg\Delta_{2}$, and that
the region of the Brillouin zone we consider satisfies $\Delta_{1}\gg k^{2}/2M$,
by dropping terms to quadratic order in $\frac{\alpha k}{\Delta_{1}-\Delta_{2}}$,
we obtain the Hamiltonian
\begin{equation}
\label{appeq:disengbeforeACstarkshift}
\left(\begin{array}{cc}
0 & -\frac{\alpha ke^{-i\theta_{k}}\text{\ensuremath{\Omega_{1}^{*}}}\text{\ensuremath{\Omega_{2}}}}{\Delta_{1}(\Delta_{2}+k^{2}/2M)}\\
-\frac{\alpha ke^{i\theta_{k}}\text{\ensuremath{\Omega_{1}}}\text{\ensuremath{\Omega_{2}^{*}}}}{\Delta_{1}(\Delta_{2}+k^{2}/2M)} & -\frac{\left|\Omega_{2}\right|^{2}}{\Delta_{2}+k^{2}/2M}
\end{array}\right).
\end{equation}
To make this approximation, we have assumed that the off-diagonal elements of Eq.\ (\ref{appeq:disengbeforeACstarkshift}) are much greater than the terms in Eq.\ (\ref{appeq:disengbeforemanyapprox}) being dropped in Eq.\ (\ref{appeq:disengbeforeACstarkshift}). There is a mismatch between the diagonal elements. To make states
$\ket{\tilde{g_{1}},\mathbf{k}}$ and $\ket{\tilde{g_{2}},\mathbf{k}}$
rotate at the same speed, we might either couple the state $\ket{\tilde{g_{1}},\mathbf{k}}$
to a band with positive curvature to induce an AC Stark shift to cancel the
first diagonal entry or couple the state $\ket{\tilde{g_{2}},\mathbf{k}}$
to some band with negative curvature to induce an AC Stark shift to
cancel the second diagonal entry. The curvatures of those auxiliary
bands have to be tuned properly during the whole process.

Now, we have engineered one term in our disentangler with $\mathcal{\mathcal{H}}\left(k\right)$
described by Eq.\ (\ref{eq:Hkthetadecom}). We can choose a different
$\Omega_{1}'$, $\Omega_{2}'$, $\Delta_{1}'$, $\Delta_{2}'$ to
generate the second term. We have to assume that the beat note between
the two schemes satisfies $\left|\Delta_{2}-\Delta'_{2}\right|\gg\left|\frac{\alpha ke^{-i\theta_{k}}\text{\ensuremath{\Omega{}_{1}^{*}}}\text{\ensuremath{\Omega{}_{2}}}}{\Delta_{1}(\Delta_{2}+k^{2}/2M)}\right|,\,\left|\frac{\alpha ke^{-i\theta_{k}}\text{\ensuremath{\Omega'{}_{1}^{*}}}\text{\ensuremath{\Omega'{}_{2}}}}{\Delta_{1}'(\Delta'_{2}+k^{2}/2M)}\right|$
to avoid crosstalk. Applying both of them at the same time, we have
the Hamiltonian in the $\ket{\tilde{g_{1}},\mathbf{k}}$, $\ket{\tilde{g_{2}},\mathbf{k}}$
basis:
\begin{equation}
\left(\begin{array}{cc}
0 & -\frac{\alpha ke^{-i\theta_{k}}\text{\ensuremath{\Omega_{1}^{*}}}\text{\ensuremath{\Omega_{2}}}}{\Delta_{1}(\Delta_{2}+k^{2}/2M)}-\frac{\alpha ke^{-i\theta_{k}}\ensuremath{\Omega'{}_{1}^{*}}\text{\ensuremath{\Omega'_{2}}}}{\Delta'_{1}(\Delta'_{2}+k^{2}/2M)}\\
-\frac{\alpha ke^{i\theta_{k}}\text{\ensuremath{\Omega_{1}}}\text{\ensuremath{\Omega_{2}^{*}}}}{\Delta_{1}(\Delta_{2}+k^{2}/2M)}-\frac{\alpha ke^{i\theta_{k}}\text{\ensuremath{\Omega'{}_{1}}}\text{\ensuremath{\Omega'{}_{2}^{*}}}}{\Delta'_{1}(\Delta'_{2}+k^{2}/2M)} & 0
\end{array}\right).
\end{equation}

Now we list all the assumptions that have been made:
\begin{enumerate}
\item The energy splittings of the dressed excited states, which are of order
$\Omega$, have to be much smaller than the hyperfine splittings of
all the states that we used. Otherwise, in FIG. \ref{appfig:engineer}(a),
we cannot use frequency selection to control each transition to engineer
synthetic selection rules.
\item All the momentum kicks should allow atoms to be in the same Brillouin
zone so that the continuum limit applies. That is, $k_{\mathrm{SOC}} \,a\ll 1$, where $a$ is the optical lattice constant.
\item $\frac{\alpha k}{\Delta_{1}-\Delta_{2}}=\frac{k_{\mathrm{SOC}}k}{M(\Delta_{1}-\Delta_{2})}\ll1$
and $\frac{k_{\mathrm{SOC}}^{2}}{M\left(\Delta_{1}-\Delta_{2}\right)}\ll1$
as well as the primed version.
\item $\Delta_{1}\gg\Delta_{2},\,k^{2}/2M$ as well as the primed version.
\item $\Delta_{1},\,\Delta_{2}\gg\Omega_{1},\:\Omega_{2}$
and $\Delta'_{1},\,\Delta'_{2}\gg\Omega'_{1},\:\Omega'_{2}$.
These two conditions imply that $\Omega\gg\Omega_{1},\:\Omega_{2},\,\Omega'_{1},\:\Omega'_{2}$.
\item The off-diagonal elements of Eq.\ (\ref{appeq:disengbeforeACstarkshift}) are much greater than the terms in Eq.\ (\ref{appeq:disengbeforemanyapprox}) being dropped in Eq.\ (\ref{appeq:disengbeforeACstarkshift}).
\item $\left|\Delta_{2}-\Delta'_{2}\right|\gg\left|\frac{\alpha ke^{-i\theta_{k}}\text{\ensuremath{\Omega{}_{1}^{*}}}\text{\ensuremath{\Omega{}_{2}}}}{\Delta_{1}(\Delta_{2}+k^{2}/2M)}\right|,\,\left|\frac{\alpha ke^{-i\theta_{k}}\text{\ensuremath{\Omega'{}_{1}^{*}}}\text{\ensuremath{\Omega'{}_{2}}}}{\Delta'_{1}(\Delta'_{2}+k^{2}/2M)}\right|$
to avoid crosstalk between the scheme determined by $\Omega_{1}$,
$\Omega_{2}$, $\Delta_{1}$, $\Delta_{2}$ and the scheme determined
by $\Omega_{1}'$, $\Omega_{2}'$, $\Delta_{1}'$, $\Delta_{2}'$.
\end{enumerate}
We remind the readers that we engineer the cMERA circuit entirely
in the interaction picture; therefore, the action of the isometry
is absorbed into that of the disentangler. The price that we have
to pay is that the disentangler is not scale-invariant at all in the
interaction picture. In principle, one can also engineer the cMERA
circuit in the Schr\"odinger picture. We leave this as a question for future research.

\section{Preparation of the Initial Near-IR State}

\label{appsec:IR}

The near-IR state with a large but finite negative $u$ is described
by Eq.\ (\ref{eq:determinedrenormalizedwavefunction}). We imagine the state to be infrared
enough that the Berry curvature is concentrated on a few momentum
points near $k=0$. Here, we describe how it can be created to use as input to the MERA circuit. A strong magnetic field should be applied to induce
hyperfine splitting in the ground-state manifold. We start with all
states in the $\ket{g_{1}}$ state, which is easy to prepare by dissipation
techniques. We then use a long-lived clock state $^{3}\mathrm{P}_{0}\;\ket{F=1/2,\:m_{F}=1/2}$
\cite{Zhang2015,Kohno2009,Lemke2009,Park2013} as a ``bus'' state
$\ket{e}$ to transfer amplitude from $\ket{g_{1}}$ to $\ket{g_{2}}$.
Seeing that $S$ states and $P$ states are well separated, we can
use a two-dimensional optical lattice to tightly trap atoms in the
$S$ states and let the atoms in the $P$ states propagate nearly
freely. We assume that the $z$ direction is tightly confined for
all states, so the corresponding degrees of freedom can be ignored.
The energy bands of $\ket{g_{1}}$ and $\ket{g_{2}}$ are flat. Here,
we assume that the $\ket{e}$ state is highly stable with a natural
linewidth much smaller than the energy splitting between the spatial
ground state and the first spatial excited state, allowing individual
momentum states to be resolved and manipulated.

In the following, we are going to use the spatial ground state of
$\ket{e}$ as a bus state. Due to open boundary conditions of optical lattices, the Bloch waves are no longer
energy eigenstates for the excited state $\ket{e}$ and we must use standing waves instead. Note
that since the eigenstates in position space of the hyperfine ground states
$\ket{g_{1}}$ and $\ket{g_{2}}$ are tightly trapped and highly degenerate,
we can still make superpositions of standing waves to create Bloch waves as energy eigenstates.
Intuitively, since particles in the hyperfine ground states $\ket{g_{1}}$
and $\ket{g_{2}}$ are tightly trapped, particles far from the boundary cannot distinguish between different boundary conditions. Our procedure to prepare the IR state
is to transfer partial amplitude from state $\ket{g_{1}}$ to $\ket{g_{2}}$
in the Brillouin zone for each $\mathbf{k}$. We denote the
lowest energy point of $\ket{e}$ as $\ket{e,0}$, which is a standing wave with small amplitude on the boundary. We couple
that state resonantly to $\ket{g_{1},\mathbf{k}}$ and $\ket{g_{2},\mathbf{k}}$
successively by different light fields, i.e., $\ket{g_{1},\mathbf{k}}\longleftrightarrow\ket{e,0}$
and then $\ket{e,0}\longleftrightarrow\ket{g_{2},\mathbf{k}}$. Other
standing waves of $\ket{e}$ are decoupled from the process due to
driving frequency mismatch. Here, we also need to ensure that other
states $\ket{g_{1},\mathbf{k}'}$ and $\ket{g_{2},\mathbf{k}'}$ with
different momenta do not interfere with the process. As a consequence,
the light fields must create a momentum selection rule for the transitions
$\ket{g_{1},\mathbf{k}}\longleftrightarrow\ket{e,0}$ and $\ket{e,0}\longleftrightarrow\ket{g_{2},\mathbf{k}}$.

We imagine a square well with wavefunction amplitude vanishing on
the periphery. This can be done by tuning the potential with spatial light modulators \cite{Fukuhara:2013aa,Nogrette2014}.
In the following, we work in the basis of the Wannier functions of the ground states and the excited state, modeling the system by a $N+2$ by $N+2$ square
lattice. We can label the lattice points by the vector $\mathbf{x}=\left(x_{1},\,x_{2}\right)$,
where $0\leq x_{1},\,x_{2}\leq N+1$, while the wavefunction vanishes
at points with $x_{1}=0,\,N+1$ or $x_{2}=0,\,N+1$. Therefore, the active
degrees of freedom for the hyperfine ground states $|g_{1}\rangle$ and $|g_{2}\rangle$
will be at $1 \leq x_{1},\,x_{2} \leq N$. In this case, the unnormalized
single-particle wavefunction of the ground state $\ket{g_{1},\mathbf{k}}$
is \cite{Busch1987}
\[
\psi_{g_{1}}(\mathbf{x})=\langle\mathbf{x}\ket{g_{1},\mathbf{k}}=\exp\left(i\,\mathbf{k}\cdot\mathbf{x}\right),
\]
where $\mathbf{k}=2\pi\left(n_{1},\,n_{2}\right)/N$ with $\:n_{1},n_{2}\in \{n\;|\;n\in\mathbb{Z}, -N/2 < n \leq N/2 \}$, and $1 \leq x_{1},\,x_{2}\leq N$. The counterpart for the excited state $|e\rangle$
is
\[
\psi_{e}(\mathbf{x})=\langle\mathbf{x}\ket{e,0}=\sin\left(\frac{\pi}{N+1}x_{1}\right)\sin\left(\frac{\pi}{N+1}x_{2}\right).
\]
Using spatial light modulators \cite{Fukuhara:2013aa,Nogrette2014}, we create the following light field:
\[
E_{g_{1}}(\mathbf{x})=\frac{\exp\left(-i\,\mathbf{q}\cdot\mathbf{x}\right)}{\sin\left(\frac{\pi}{N+1}x_{1}\right)\sin\left(\frac{\pi}{N+1}x_{2}\right)},
\]
where $\mathbf{q}=2\pi\left(m_{1},\,m_{2}\right)/N,\:m_{1},m_{2}\in\mathbb{Z}$.
A momentum selection rule for $\ket{g_{1},\mathbf{k}}\longleftrightarrow\ket{e,0}$
can now be engineered:
\begin{align}
\sum_{\mathbf{x}}\psi_{e}(\mathbf{x})\,E_{g_{1}}(\mathbf{x})\,\psi_{g_{1}}(\mathbf{x}) & =\sum_{\mathbf{x}}\sin\left(\frac{\pi}{N+1}x_{1}\right)\sin\left(\frac{\pi}{N+1}x_{2}\right)\,\frac{\exp\left(-i\,\mathbf{q}\cdot\mathbf{x}\right)}{\sin\left(\frac{\pi}{N+1}x\right)\sin\left(\frac{\pi}{N+1}x_{2}\right)}\,\exp\left(i\,\mathbf{k}\cdot\mathbf{x}\right)\nonumber \\ 
 & =\sum_{\mathbf{x}}\exp\left(i\,(\mathbf{k}-\mathbf{q})\cdot\mathbf{x}\right)\propto\delta_{\mathbf{k},\mathbf{q}}.
\end{align}
Notice that since the points where the denominator of $E(\mathbf{x})$
becomes zero are excluded from our consideration, the light field
is well defined. A similar selection
rule can be derived for $\ket{e,0}\longleftrightarrow\ket{g_{2},\mathbf{k}}$.

With this technique in mind, we can adjust the relative amplitude
between $\ket{g_{1}}$ and $\ket{g_{2}}$ in the Brillouin zone to
create the near-IR state described in Eq.\ (\ref{eq:determinedrenormalizedwavefunction})
by fine-tuning phases and durations of the light field pulses. Given
that the Berry curvature is concentrated on a few momentum points near
$k=0$, we can limit this procedure to only a few small momentum points without too much error.

\begin{thebibliography}{60}%
\makeatletter
\providecommand \@ifxundefined [1]{%
 \@ifx{#1\undefined}
}%
\providecommand \@ifnum [1]{%
 \ifnum #1\expandafter \@firstoftwo
 \else \expandafter \@secondoftwo
 \fi
}%
\providecommand \@ifx [1]{%
 \ifx #1\expandafter \@firstoftwo
 \else \expandafter \@secondoftwo
 \fi
}%
\providecommand \natexlab [1]{#1}%
\providecommand \enquote  [1]{``#1''}%
\providecommand \bibnamefont  [1]{#1}%
\providecommand \bibfnamefont [1]{#1}%
\providecommand \citenamefont [1]{#1}%
\providecommand \href@noop [0]{\@secondoftwo}%
\providecommand \href [0]{\begingroup \@sanitize@url \@href}%
\providecommand \@href[1]{\@@startlink{#1}\@@href}%
\providecommand \@@href[1]{\endgroup#1\@@endlink}%
\providecommand \@sanitize@url [0]{\catcode `\\12\catcode `\$12\catcode
  `\&12\catcode `\#12\catcode `\^12\catcode `\_12\catcode `\%12\relax}%
\providecommand \@@startlink[1]{}%
\providecommand \@@endlink[0]{}%
\providecommand \url  [0]{\begingroup\@sanitize@url \@url }%
\providecommand \@url [1]{\endgroup\@href {#1}{\urlprefix }}%
\providecommand \urlprefix  [0]{URL }%
\providecommand \Eprint [0]{\href }%
\providecommand \doibase [0]{http://dx.doi.org/}%
\providecommand \selectlanguage [0]{\@gobble}%
\providecommand \bibinfo  [0]{\@secondoftwo}%
\providecommand \bibfield  [0]{\@secondoftwo}%
\providecommand \translation [1]{[#1]}%
\providecommand \BibitemOpen [0]{}%
\providecommand \bibitemStop [0]{}%
\providecommand \bibitemNoStop [0]{.\EOS\space}%
\providecommand \EOS [0]{\spacefactor3000\relax}%
\providecommand \BibitemShut  [1]{\csname bibitem#1\endcsname}%
\let\auto@bib@innerbib\@empty
\bibitem [{\citenamefont {{Or{\'u}s}}(2014)}]{Orus2014}%
  \BibitemOpen
  \bibfield  {author} {\bibinfo {author} {\bibfnamefont {R.}~\bibnamefont
  {{Or{\'u}s}}},\ }\bibfield  {title} {\enquote {\bibinfo {title} {{A practical
  introduction to tensor networks: Matrix product states and projected
  entangled pair states}},}\ }\href@noop {} {\bibfield  {journal} {\bibinfo
  {journal} {Ann. Phys. (N.Y.)}\ }\textbf {\bibinfo {volume} {349}},\ \bibinfo
  {pages} {117--158} (\bibinfo {year} {2014})}\BibitemShut {NoStop}%
\bibitem [{\citenamefont {{Bridgeman}}\ and\ \citenamefont
  {{Chubb}}(2017)}]{Bridgeman2016}%
  \BibitemOpen
  \bibfield  {author} {\bibinfo {author} {\bibfnamefont {J.~C.}\ \bibnamefont
  {{Bridgeman}}}\ and\ \bibinfo {author} {\bibfnamefont {C.~T.}\ \bibnamefont
  {{Chubb}}},\ }\bibfield  {title} {\enquote {\bibinfo {title} {{Hand-waving
  and interpretive dance: an introductory course on tensor networks}},}\
  }\href@noop {} {\bibfield  {journal} {\bibinfo  {journal} {J. Phys. A}\
  }\textbf {\bibinfo {volume} {50}},\ \bibinfo {eid} {223001} (\bibinfo {year}
  {2017})}\BibitemShut {NoStop}%
\bibitem [{\citenamefont {{Vidal}}(2008)}]{Vidal2006}%
  \BibitemOpen
  \bibfield  {author} {\bibinfo {author} {\bibfnamefont {G.}~\bibnamefont
  {{Vidal}}},\ }\bibfield  {title} {\enquote {\bibinfo {title} {{Class of
  Quantum Many-Body States That Can Be Efficiently Simulated}},}\ }\href@noop
  {} {\bibfield  {journal} {\bibinfo  {journal} {\prl}\ }\textbf {\bibinfo
  {volume} {101}},\ \bibinfo {eid} {110501} (\bibinfo {year}
  {2008})}\BibitemShut {NoStop}%
\bibitem [{\citenamefont {Verstraete}\ and\ \citenamefont
  {Cirac}(2006)}]{Verstraete2006}%
  \BibitemOpen
  \bibfield  {author} {\bibinfo {author} {\bibfnamefont {F.}~\bibnamefont
  {Verstraete}}\ and\ \bibinfo {author} {\bibfnamefont {J.~I.}\ \bibnamefont
  {Cirac}},\ }\bibfield  {title} {\enquote {\bibinfo {title} {{Matrix product
  states represent ground states faithfully}},}\ }\href@noop {} {\bibfield
  {journal} {\bibinfo  {journal} {\prb}\ }\textbf {\bibinfo {volume} {73}},\
  \bibinfo {pages} {094423} (\bibinfo {year} {2006})}\BibitemShut {NoStop}%
\bibitem [{\citenamefont {Verstraete}\ \emph {et~al.}(2006)\citenamefont
  {Verstraete}, \citenamefont {Wolf}, \citenamefont {Perez-Garcia},\ and\
  \citenamefont {Cirac}}]{Verstraete2006a}%
  \BibitemOpen
  \bibfield  {author} {\bibinfo {author} {\bibfnamefont {F.}~\bibnamefont
  {Verstraete}}, \bibinfo {author} {\bibfnamefont {M.~M.}\ \bibnamefont
  {Wolf}}, \bibinfo {author} {\bibfnamefont {D.}~\bibnamefont {Perez-Garcia}},
  \ and\ \bibinfo {author} {\bibfnamefont {J.~I.}\ \bibnamefont {Cirac}},\
  }\bibfield  {title} {\enquote {\bibinfo {title} {{Criticality, the area law,
  and the computational power of projected entangled pair states}},}\
  }\href@noop {} {\bibfield  {journal} {\bibinfo  {journal} {\prl}\ }\textbf
  {\bibinfo {volume} {96}},\ \bibinfo {pages} {220601} (\bibinfo {year}
  {2006})}\BibitemShut {NoStop}%
\bibitem [{\citenamefont {Hastings}(2007)}]{Hastings2007}%
  \BibitemOpen
  \bibfield  {author} {\bibinfo {author} {\bibfnamefont {M.~B.}\ \bibnamefont
  {Hastings}},\ }\bibfield  {title} {\enquote {\bibinfo {title} {{An area law
  for one-dimensional quantum systems}},}\ }\href@noop {} {\bibfield  {journal}
  {\bibinfo  {journal} {{J.\ Stat.\ Mech.:\ Theor.\ Exp.}}\ }\textbf {\bibinfo
  {volume} {2007}},\ \bibinfo {pages} {P08024} (\bibinfo {year}
  {2007})}\BibitemShut {NoStop}%
\bibitem [{\citenamefont {Vidal}(2007)}]{Vidal2007}%
  \BibitemOpen
  \bibfield  {author} {\bibinfo {author} {\bibfnamefont {G.}~\bibnamefont
  {Vidal}},\ }\bibfield  {title} {\enquote {\bibinfo {title} {{Entanglement
  renormalization}},}\ }\href@noop {} {\bibfield  {journal} {\bibinfo
  {journal} {\prl}\ }\textbf {\bibinfo {volume} {99}},\ \bibinfo {pages}
  {220405} (\bibinfo {year} {2007})}\BibitemShut {NoStop}%
\bibitem [{\citenamefont {{Wolf}}\ \emph {et~al.}(2008)\citenamefont {{Wolf}},
  \citenamefont {{Verstraete}}, \citenamefont {{Hastings}},\ and\ \citenamefont
  {{Cirac}}}]{Wolf2008}%
  \BibitemOpen
  \bibfield  {author} {\bibinfo {author} {\bibfnamefont {M.~M.}\ \bibnamefont
  {{Wolf}}}, \bibinfo {author} {\bibfnamefont {F.}~\bibnamefont
  {{Verstraete}}}, \bibinfo {author} {\bibfnamefont {M.~B.}\ \bibnamefont
  {{Hastings}}}, \ and\ \bibinfo {author} {\bibfnamefont {J.~I.}\ \bibnamefont
  {{Cirac}}},\ }\bibfield  {title} {\enquote {\bibinfo {title} {{Area Laws in
  Quantum Systems: Mutual Information and Correlations}},}\ }\href@noop {}
  {\bibfield  {journal} {\bibinfo  {journal} {\prl}\ }\textbf {\bibinfo
  {volume} {100}},\ \bibinfo {eid} {070502} (\bibinfo {year}
  {2008})}\BibitemShut {NoStop}%
\bibitem [{\citenamefont {Wilson}(1974)}]{Wilson1974}%
  \BibitemOpen
  \bibfield  {author} {\bibinfo {author} {\bibfnamefont {K.~G.}\ \bibnamefont
  {Wilson}},\ }\bibfield  {title} {\enquote {\bibinfo {title} {{The
  renormalization group and the $\epsilon$ expansion}},}\ }\href@noop {}
  {\bibfield  {journal} {\bibinfo  {journal} {Phys. Rep.}\ }\textbf {\bibinfo
  {volume} {12}},\ \bibinfo {pages} {75--199} (\bibinfo {year}
  {1974})}\BibitemShut {NoStop}%
\bibitem [{\citenamefont {Wilson}(1975)}]{Wilson1975}%
  \BibitemOpen
  \bibfield  {author} {\bibinfo {author} {\bibfnamefont {K.~G.}\ \bibnamefont
  {Wilson}},\ }\bibfield  {title} {\enquote {\bibinfo {title} {{The
  renormalization group: Critical phenomena and the Kondo problem}},}\
  }\href@noop {} {\bibfield  {journal} {\bibinfo  {journal} {Rev. Mod. Phys.}\
  }\textbf {\bibinfo {volume} {47}},\ \bibinfo {pages} {773--840} (\bibinfo
  {year} {1975})}\BibitemShut {NoStop}%
\bibitem [{\citenamefont {Zinn-Justin}(2007)}]{Zinn-Justin2007}%
  \BibitemOpen
  \bibfield  {author} {\bibinfo {author} {\bibfnamefont {J.}~\bibnamefont
  {Zinn-Justin}},\ }\href@noop {} {\emph {\bibinfo {title} {{Phase transitions
  and renormalization group}}}}\ (\bibinfo  {publisher} {Oxford},\ \bibinfo
  {year} {2007})\BibitemShut {NoStop}%
\bibitem [{\citenamefont {Aguado}\ and\ \citenamefont
  {Vidal}(2008)}]{Aguado2008}%
  \BibitemOpen
  \bibfield  {author} {\bibinfo {author} {\bibfnamefont {M.}~\bibnamefont
  {Aguado}}\ and\ \bibinfo {author} {\bibfnamefont {G.}~\bibnamefont {Vidal}},\
  }\bibfield  {title} {\enquote {\bibinfo {title} {{Entanglement
  renormalization and topological order}},}\ }\href@noop {} {\bibfield
  {journal} {\bibinfo  {journal} {\prl}\ }\textbf {\bibinfo {volume} {100}},\
  \bibinfo {pages} {070404} (\bibinfo {year} {2008})}\BibitemShut {NoStop}%
\bibitem [{\citenamefont {Pfeifer}\ \emph {et~al.}(2009)\citenamefont
  {Pfeifer}, \citenamefont {Evenbly},\ and\ \citenamefont
  {Vidal}}]{Pfeifer2009}%
  \BibitemOpen
  \bibfield  {author} {\bibinfo {author} {\bibfnamefont {R.~N.~C.}\
  \bibnamefont {Pfeifer}}, \bibinfo {author} {\bibfnamefont {G.}~\bibnamefont
  {Evenbly}}, \ and\ \bibinfo {author} {\bibfnamefont {G.}~\bibnamefont
  {Vidal}},\ }\bibfield  {title} {\enquote {\bibinfo {title} {{Entanglement
  renormalization, scale invariance, and quantum criticality}},}\ }\href@noop
  {} {\bibfield  {journal} {\bibinfo  {journal} {\pra}\ }\textbf {\bibinfo
  {volume} {79}},\ \bibinfo {pages} {040301} (\bibinfo {year}
  {2009})}\BibitemShut {NoStop}%
\bibitem [{\citenamefont {K{\"{o}}nig}\ \emph {et~al.}(2009)\citenamefont
  {K{\"{o}}nig}, \citenamefont {Reichardt},\ and\ \citenamefont
  {Vidal}}]{Konig2009}%
  \BibitemOpen
  \bibfield  {author} {\bibinfo {author} {\bibfnamefont {R.}~\bibnamefont
  {K{\"{o}}nig}}, \bibinfo {author} {\bibfnamefont {B.~W.}\ \bibnamefont
  {Reichardt}}, \ and\ \bibinfo {author} {\bibfnamefont {G.}~\bibnamefont
  {Vidal}},\ }\bibfield  {title} {\enquote {\bibinfo {title} {{Exact
  entanglement renormalization for string-net models}},}\ }\href@noop {}
  {\bibfield  {journal} {\bibinfo  {journal} {\prb}\ }\textbf {\bibinfo
  {volume} {79}},\ \bibinfo {pages} {195123} (\bibinfo {year}
  {2009})}\BibitemShut {NoStop}%
\bibitem [{\citenamefont {K{\"{o}}nig}\ and\ \citenamefont
  {Bilgin}(2010)}]{Konig2010}%
  \BibitemOpen
  \bibfield  {author} {\bibinfo {author} {\bibfnamefont {R.}~\bibnamefont
  {K{\"{o}}nig}}\ and\ \bibinfo {author} {\bibfnamefont {E.}~\bibnamefont
  {Bilgin}},\ }\bibfield  {title} {\enquote {\bibinfo {title} {{Anyonic
  entanglement renormalization}},}\ }\href@noop {} {\bibfield  {journal}
  {\bibinfo  {journal} {\prb}\ }\textbf {\bibinfo {volume} {82}},\ \bibinfo
  {pages} {125118} (\bibinfo {year} {2010})}\BibitemShut {NoStop}%
\bibitem [{\citenamefont {{Singh}}\ and\ \citenamefont
  {{Vidal}}(2013)}]{Singh2013}%
  \BibitemOpen
  \bibfield  {author} {\bibinfo {author} {\bibfnamefont {S.}~\bibnamefont
  {{Singh}}}\ and\ \bibinfo {author} {\bibfnamefont {G.}~\bibnamefont
  {{Vidal}}},\ }\bibfield  {title} {\enquote {\bibinfo {title}
  {{Symmetry-protected entanglement renormalization}},}\ }\href@noop {}
  {\bibfield  {journal} {\bibinfo  {journal} {\prb}\ }\textbf {\bibinfo
  {volume} {88}},\ \bibinfo {pages} {121108} (\bibinfo {year}
  {2013})}\BibitemShut {NoStop}%
\bibitem [{\citenamefont {{Evenbly}}\ and\ \citenamefont
  {{White}}(2016)}]{Evenbly2016}%
  \BibitemOpen
  \bibfield  {author} {\bibinfo {author} {\bibfnamefont {G.}~\bibnamefont
  {{Evenbly}}}\ and\ \bibinfo {author} {\bibfnamefont {S.~R.}\ \bibnamefont
  {{White}}},\ }\bibfield  {title} {\enquote {\bibinfo {title} {{Entanglement
  Renormalization and Wavelets}},}\ }\href@noop {} {\bibfield  {journal}
  {\bibinfo  {journal} {\prl}\ }\textbf {\bibinfo {volume} {116}},\ \bibinfo
  {pages} {140403} (\bibinfo {year} {2016})}\BibitemShut {NoStop}%
\bibitem [{\citenamefont {{Haegeman}}\ \emph {et~al.}(2018)\citenamefont
  {{Haegeman}}, \citenamefont {{Swingle}}, \citenamefont {{Walter}},
  \citenamefont {{Cotler}}, \citenamefont {{Evenbly}},\ and\ \citenamefont
  {{Scholz}}}]{Haegeman2017}%
  \BibitemOpen
  \bibfield  {author} {\bibinfo {author} {\bibfnamefont {J.}~\bibnamefont
  {{Haegeman}}}, \bibinfo {author} {\bibfnamefont {B.}~\bibnamefont
  {{Swingle}}}, \bibinfo {author} {\bibfnamefont {M.}~\bibnamefont {{Walter}}},
  \bibinfo {author} {\bibfnamefont {J.}~\bibnamefont {{Cotler}}}, \bibinfo
  {author} {\bibfnamefont {G.}~\bibnamefont {{Evenbly}}}, \ and\ \bibinfo
  {author} {\bibfnamefont {V.~B.}\ \bibnamefont {{Scholz}}},\ }\bibfield
  {title} {\enquote {\bibinfo {title} {{Rigorous Free-Fermion Entanglement
  Renormalization from Wavelet Theory}},}\ }\href@noop {} {\bibfield  {journal}
  {\bibinfo  {journal} {Phys. Rev. X}\ }\textbf {\bibinfo {volume} {8}},\
  \bibinfo {eid} {011003} (\bibinfo {year} {2018})}\BibitemShut {NoStop}%
\bibitem [{\citenamefont {{Hansson}}\ \emph {et~al.}(2017)\citenamefont
  {{Hansson}}, \citenamefont {{Hermanns}}, \citenamefont {{Simon}},\ and\
  \citenamefont {{Viefers}}}]{Hansson2016}%
  \BibitemOpen
  \bibfield  {author} {\bibinfo {author} {\bibfnamefont {T.~H.}\ \bibnamefont
  {{Hansson}}}, \bibinfo {author} {\bibfnamefont {M.}~\bibnamefont
  {{Hermanns}}}, \bibinfo {author} {\bibfnamefont {S.~H.}\ \bibnamefont
  {{Simon}}}, \ and\ \bibinfo {author} {\bibfnamefont {S.~F.}\ \bibnamefont
  {{Viefers}}},\ }\bibfield  {title} {\enquote {\bibinfo {title} {{Quantum Hall
  physics: Hierarchies and conformal field theory techniques}},}\ }\href@noop
  {} {\bibfield  {journal} {\bibinfo  {journal} {\rmp}\ }\textbf {\bibinfo
  {volume} {89}},\ \bibinfo {eid} {025005} (\bibinfo {year}
  {2017})}\BibitemShut {NoStop}%
\bibitem [{\citenamefont {{Wen}}(2017)}]{Wen2016a}%
  \BibitemOpen
  \bibfield  {author} {\bibinfo {author} {\bibfnamefont {X.-G.}\ \bibnamefont
  {{Wen}}},\ }\bibfield  {title} {\enquote {\bibinfo {title} {{Colloquium: Zoo
  of quantum-topological phases of matter}},}\ }\href@noop {} {\bibfield
  {journal} {\bibinfo  {journal} {\rmp}\ }\textbf {\bibinfo {volume} {89}},\
  \bibinfo {pages} {041004} (\bibinfo {year} {2017})}\BibitemShut {NoStop}%
\bibitem [{\citenamefont {Nayak}\ \emph {et~al.}(2008)\citenamefont {Nayak},
  \citenamefont {Simon}, \citenamefont {Stern}, \citenamefont {Freedman},\ and\
  \citenamefont {{Das Sarma}}}]{Nayak2008}%
  \BibitemOpen
  \bibfield  {author} {\bibinfo {author} {\bibfnamefont {C.}~\bibnamefont
  {Nayak}}, \bibinfo {author} {\bibfnamefont {S.~H.}\ \bibnamefont {Simon}},
  \bibinfo {author} {\bibfnamefont {A.}~\bibnamefont {Stern}}, \bibinfo
  {author} {\bibfnamefont {M.}~\bibnamefont {Freedman}}, \ and\ \bibinfo
  {author} {\bibfnamefont {S.}~\bibnamefont {{Das Sarma}}},\ }\bibfield
  {title} {\enquote {\bibinfo {title} {{Non-Abelian anyons and topological
  quantum computation}},}\ }\href@noop {} {\bibfield  {journal} {\bibinfo
  {journal} {\rmp}\ }\textbf {\bibinfo {volume} {80}},\ \bibinfo {pages}
  {1083--1159} (\bibinfo {year} {2008})}\BibitemShut {NoStop}%
\bibitem [{\citenamefont {{Bravyi}}\ \emph {et~al.}(2006)\citenamefont
  {{Bravyi}}, \citenamefont {{Hastings}},\ and\ \citenamefont
  {{Verstraete}}}]{Bravyi2006}%
  \BibitemOpen
  \bibfield  {author} {\bibinfo {author} {\bibfnamefont {S.}~\bibnamefont
  {{Bravyi}}}, \bibinfo {author} {\bibfnamefont {M.~B.}\ \bibnamefont
  {{Hastings}}}, \ and\ \bibinfo {author} {\bibfnamefont {F.}~\bibnamefont
  {{Verstraete}}},\ }\bibfield  {title} {\enquote {\bibinfo {title}
  {{Lieb-Robinson Bounds and the Generation of Correlations and Topological
  Quantum Order}},}\ }\href@noop {} {\bibfield  {journal} {\bibinfo  {journal}
  {\prl}\ }\textbf {\bibinfo {volume} {97}},\ \bibinfo {eid} {050401} (\bibinfo
  {year} {2006})}\BibitemShut {NoStop}%
\bibitem [{\citenamefont {Barthel}\ \emph {et~al.}(2010)\citenamefont
  {Barthel}, \citenamefont {Kliesch},\ and\ \citenamefont
  {Eisert}}]{Barthel2010}%
  \BibitemOpen
  \bibfield  {author} {\bibinfo {author} {\bibfnamefont {T.}~\bibnamefont
  {Barthel}}, \bibinfo {author} {\bibfnamefont {M.}~\bibnamefont {Kliesch}}, \
  and\ \bibinfo {author} {\bibfnamefont {J.}~\bibnamefont {Eisert}},\
  }\bibfield  {title} {\enquote {\bibinfo {title} {{Real-space renormalization
  yields finite correlations}},}\ }\href@noop {} {\bibfield  {journal}
  {\bibinfo  {journal} {\prl}\ }\textbf {\bibinfo {volume} {105}},\ \bibinfo
  {pages} {010502} (\bibinfo {year} {2010})}\BibitemShut {NoStop}%
\bibitem [{\citenamefont {Dubail}\ and\ \citenamefont
  {Read}(2015)}]{Dubail2015}%
  \BibitemOpen
  \bibfield  {author} {\bibinfo {author} {\bibfnamefont {J.}~\bibnamefont
  {Dubail}}\ and\ \bibinfo {author} {\bibfnamefont {N.}~\bibnamefont {Read}},\
  }\bibfield  {title} {\enquote {\bibinfo {title} {{Tensor network trial states
  for chiral topological phases in two dimensions and a no-go theorem in any
  dimension}},}\ }\href@noop {} {\bibfield  {journal} {\bibinfo  {journal}
  {\prb}\ }\textbf {\bibinfo {volume} {92}},\ \bibinfo {pages} {205307}
  (\bibinfo {year} {2015})}\BibitemShut {NoStop}%
\bibitem [{\citenamefont {Wahl}\ \emph {et~al.}(2013)\citenamefont {Wahl},
  \citenamefont {Tu}, \citenamefont {Schuch},\ and\ \citenamefont
  {Cirac}}]{Wahl2013}%
  \BibitemOpen
  \bibfield  {author} {\bibinfo {author} {\bibfnamefont {T.~B.}\ \bibnamefont
  {Wahl}}, \bibinfo {author} {\bibfnamefont {H.~H.}\ \bibnamefont {Tu}},
  \bibinfo {author} {\bibfnamefont {N.}~\bibnamefont {Schuch}}, \ and\ \bibinfo
  {author} {\bibfnamefont {J.~I.}\ \bibnamefont {Cirac}},\ }\bibfield  {title}
  {\enquote {\bibinfo {title} {{Projected entangled-pair states can describe
  chiral topological states}},}\ }\href@noop {} {\bibfield  {journal} {\bibinfo
   {journal} {\prl}\ }\textbf {\bibinfo {volume} {111}},\ \bibinfo {pages}
  {236805} (\bibinfo {year} {2013})}\BibitemShut {NoStop}%
\bibitem [{\citenamefont {{Li}}\ and\ \citenamefont {{Mong}}(2017)}]{Li2017}%
  \BibitemOpen
  \bibfield  {author} {\bibinfo {author} {\bibfnamefont {Z.}~\bibnamefont
  {{Li}}}\ and\ \bibinfo {author} {\bibfnamefont {R.~S.~K.}\ \bibnamefont
  {{Mong}}},\ }\bibfield  {title} {\enquote {\bibinfo {title} {{Entanglement
  renormalization for chiral topological phases}},}\ }\href@noop ,\ \Eprint {http://arxiv.org/abs/1703.00464}
  {arXiv:1703.00464 (2017)} \BibitemShut {NoStop}%
\bibitem [{\citenamefont {Haegeman}\ \emph {et~al.}(2013)\citenamefont
  {Haegeman}, \citenamefont {Osborne}, \citenamefont {Verschelde},\ and\
  \citenamefont {Verstraete}}]{Haegeman2013}%
  \BibitemOpen
  \bibfield  {author} {\bibinfo {author} {\bibfnamefont {J.}~\bibnamefont
  {Haegeman}}, \bibinfo {author} {\bibfnamefont {T.~J.}\ \bibnamefont
  {Osborne}}, \bibinfo {author} {\bibfnamefont {H.}~\bibnamefont {Verschelde}},
  \ and\ \bibinfo {author} {\bibfnamefont {F.}~\bibnamefont {Verstraete}},\
  }\bibfield  {title} {\enquote {\bibinfo {title} {{Entanglement
  renormalization for quantum fields in real space}},}\ }\href@noop {}
  {\bibfield  {journal} {\bibinfo  {journal} {\prl}\ }\textbf {\bibinfo
  {volume} {110}},\ \bibinfo {pages} {100402} (\bibinfo {year}
  {2013})}\BibitemShut {NoStop}%
\bibitem [{\citenamefont {Hu}\ and\ \citenamefont {Vidal}(2017)}]{Hu2017}%
  \BibitemOpen
  \bibfield  {author} {\bibinfo {author} {\bibfnamefont {Q.}~\bibnamefont
  {Hu}}\ and\ \bibinfo {author} {\bibfnamefont {G.}~\bibnamefont {Vidal}},\
  }\bibfield  {title} {\enquote {\bibinfo {title} {{Spacetime Symmetries and
  Conformal Data in the Continuous Multiscale Entanglement Renormalization
  Ansatz}},}\ }\href@noop {} {\bibfield  {journal} {\bibinfo  {journal} {\prl}\
  }\textbf {\bibinfo {volume} {119}},\ \bibinfo {pages} {010603} (\bibinfo
  {year} {2017})}\BibitemShut {NoStop}%
\bibitem [{\citenamefont {{Bernevig}}\ \emph {et~al.}(2006)\citenamefont
  {{Bernevig}}, \citenamefont {{Hughes}},\ and\ \citenamefont
  {{Zhang}}}]{Bernevig2006a}%
  \BibitemOpen
  \bibfield  {author} {\bibinfo {author} {\bibfnamefont {B.~A.}\ \bibnamefont
  {{Bernevig}}}, \bibinfo {author} {\bibfnamefont {T.~L.}\ \bibnamefont
  {{Hughes}}}, \ and\ \bibinfo {author} {\bibfnamefont {S.-C.}\ \bibnamefont
  {{Zhang}}},\ }\bibfield  {title} {\enquote {\bibinfo {title} {{Quantum Spin
  Hall Effect and Topological Phase Transition in HgTe Quantum Wells}},}\
  }\href@noop {} {\bibfield  {journal} {\bibinfo  {journal} {Science}\ }\textbf
  {\bibinfo {volume} {314}},\ \bibinfo {pages} {1757} (\bibinfo {year}
  {2006})}\BibitemShut {NoStop}%
\bibitem [{\citenamefont {Wen}\ \emph {et~al.}(2016)\citenamefont {Wen},
  \citenamefont {Cho}, \citenamefont {Lopes}, \citenamefont {Gu}, \citenamefont
  {Qi},\ and\ \citenamefont {Ryu}}]{Wen2016}%
  \BibitemOpen
  \bibfield  {author} {\bibinfo {author} {\bibfnamefont {X.}~\bibnamefont
  {Wen}}, \bibinfo {author} {\bibfnamefont {G.~Y.}\ \bibnamefont {Cho}},
  \bibinfo {author} {\bibfnamefont {P.~L.S.}\ \bibnamefont {Lopes}}, \bibinfo
  {author} {\bibfnamefont {Y.}~\bibnamefont {Gu}}, \bibinfo {author}
  {\bibfnamefont {X.~L.}\ \bibnamefont {Qi}}, \ and\ \bibinfo {author}
  {\bibfnamefont {S.}~\bibnamefont {Ryu}},\ }\bibfield  {title} {\enquote
  {\bibinfo {title} {{Holographic entanglement renormalization of topological
  insulators}},}\ }\href@noop {} {\bibfield  {journal} {\bibinfo  {journal}
  {\prb}\ }\textbf {\bibinfo {volume} {94}},\ \bibinfo {pages} {075124}
  (\bibinfo {year} {2016})}\BibitemShut {NoStop}%
\bibitem [{\citenamefont {Swingle}\ and\ \citenamefont
  {McGreevy}(2016)}]{Swingle2016}%
  \BibitemOpen
  \bibfield  {author} {\bibinfo {author} {\bibfnamefont {B.}~\bibnamefont
  {Swingle}}\ and\ \bibinfo {author} {\bibfnamefont {J.}~\bibnamefont
  {McGreevy}},\ }\bibfield  {title} {\enquote {\bibinfo {title}
  {{Renormalization group constructions of topological quantum liquids and
  beyond}},}\ }\href@noop {} {\bibfield  {journal} {\bibinfo  {journal} {\prb}\
  }\textbf {\bibinfo {volume} {93}},\ \bibinfo {pages} {045127} (\bibinfo
  {year} {2016})}\BibitemShut {NoStop}%
\bibitem [{\citenamefont {{Huang}}\ and\ \citenamefont
  {{Moore}}(2017)}]{Huang2017}%
  \BibitemOpen
  \bibfield  {author} {\bibinfo {author} {\bibfnamefont {Y.}~\bibnamefont
  {{Huang}}}\ and\ \bibinfo {author} {\bibfnamefont {J.~E.}\ \bibnamefont
  {{Moore}}},\ }\bibfield  {title} {\enquote {\bibinfo {title} {{Neural network
  representation of tensor network and chiral states}},}\ }\href@noop \ \Eprint {http://arxiv.org/abs/1701.06246}
  {arXiv:1701.06246 (2017)} \BibitemShut {NoStop}%
\bibitem [{\citenamefont {{Kaubruegger}}\ \emph {et~al.}(2018)\citenamefont
  {{Kaubruegger}}, \citenamefont {{Pastori}},\ and\ \citenamefont
  {{Budich}}}]{Kaubruegger2017}%
  \BibitemOpen
  \bibfield  {author} {\bibinfo {author} {\bibfnamefont {R.}~\bibnamefont
  {{Kaubruegger}}}, \bibinfo {author} {\bibfnamefont {L.}~\bibnamefont
  {{Pastori}}}, \ and\ \bibinfo {author} {\bibfnamefont {J.~C.}\ \bibnamefont
  {{Budich}}},\ }\bibfield  {title} {\enquote {\bibinfo {title} {{Chiral
  topological phases from artificial neural networks}},}\ }\href@noop {}
  {\bibfield  {journal} {\bibinfo  {journal} {\prb}\ }\textbf {\bibinfo
  {volume} {97}},\ \bibinfo {eid} {195136} (\bibinfo {year}
  {2018})}\BibitemShut {NoStop}%
\bibitem [{\citenamefont {{Glasser}}\ \emph {et~al.}(2018)\citenamefont
  {{Glasser}}, \citenamefont {{Pancotti}}, \citenamefont {{August}},
  \citenamefont {{Rodriguez}},\ and\ \citenamefont {{Cirac}}}]{Glasser2017}%
  \BibitemOpen
  \bibfield  {author} {\bibinfo {author} {\bibfnamefont {I.}~\bibnamefont
  {{Glasser}}}, \bibinfo {author} {\bibfnamefont {N.}~\bibnamefont
  {{Pancotti}}}, \bibinfo {author} {\bibfnamefont {M.}~\bibnamefont
  {{August}}}, \bibinfo {author} {\bibfnamefont {I.~D.}\ \bibnamefont
  {{Rodriguez}}}, \ and\ \bibinfo {author} {\bibfnamefont {J.~I.}\ \bibnamefont
  {{Cirac}}},\ }\bibfield  {title} {\enquote {\bibinfo {title} {{Neural-Network
  Quantum States, String-Bond States, and Chiral Topological States}},}\
  }\href@noop {} {\bibfield  {journal} {\bibinfo  {journal} {Phys. Rev. X}\
  }\textbf {\bibinfo {volume} {8}},\ \bibinfo {pages} {011006} (\bibinfo {year}
  {2018})}\BibitemShut {NoStop}%
\bibitem [{\citenamefont {Wahl}\ \emph {et~al.}(2014)\citenamefont {Wahl},
  \citenamefont {Ha{\ss}ler}, \citenamefont {Tu}, \citenamefont {Cirac},\ and\
  \citenamefont {Schuch}}]{Wahl2014}%
  \BibitemOpen
  \bibfield  {author} {\bibinfo {author} {\bibfnamefont {T.~B.}\ \bibnamefont
  {Wahl}}, \bibinfo {author} {\bibfnamefont {S.~T.}\ \bibnamefont
  {Ha{\ss}ler}}, \bibinfo {author} {\bibfnamefont {H.~H.}\ \bibnamefont {Tu}},
  \bibinfo {author} {\bibfnamefont {J.~I.}\ \bibnamefont {Cirac}}, \ and\
  \bibinfo {author} {\bibfnamefont {N.}~\bibnamefont {Schuch}},\ }\bibfield
  {title} {\enquote {\bibinfo {title} {{Symmetries and boundary theories for
  chiral projected entangled pair states}},}\ }\href@noop {} {\bibfield
  {journal} {\bibinfo  {journal} {\prb}\ }\textbf {\bibinfo {volume} {90}},\
  \bibinfo {pages} {115133} (\bibinfo {year} {2014})}\BibitemShut {NoStop}%
\bibitem [{\citenamefont {Poilblanc}\ \emph {et~al.}(2015)\citenamefont
  {Poilblanc}, \citenamefont {Cirac},\ and\ \citenamefont
  {Schuch}}]{Poilblanc2015}%
  \BibitemOpen
  \bibfield  {author} {\bibinfo {author} {\bibfnamefont {D.}~\bibnamefont
  {Poilblanc}}, \bibinfo {author} {\bibfnamefont {J.~I.}\ \bibnamefont
  {Cirac}}, \ and\ \bibinfo {author} {\bibfnamefont {N.}~\bibnamefont
  {Schuch}},\ }\bibfield  {title} {\enquote {\bibinfo {title} {{Chiral
  topological spin liquids with projected entangled pair states}},}\
  }\href@noop {} {\bibfield  {journal} {\bibinfo  {journal} {\prb}\ }\textbf
  {\bibinfo {volume} {91}},\ \bibinfo {pages} {224431} (\bibinfo {year}
  {2015})}\BibitemShut {NoStop}%
\bibitem [{\citenamefont {Poilblanc}\ \emph {et~al.}(2016)\citenamefont
  {Poilblanc}, \citenamefont {Schuch},\ and\ \citenamefont
  {Affleck}}]{Poilblanc2016}%
  \BibitemOpen
  \bibfield  {author} {\bibinfo {author} {\bibfnamefont {D.}~\bibnamefont
  {Poilblanc}}, \bibinfo {author} {\bibfnamefont {N.}~\bibnamefont {Schuch}}, \
  and\ \bibinfo {author} {\bibfnamefont {I.}~\bibnamefont {Affleck}},\
  }\bibfield  {title} {\enquote {\bibinfo {title} {{SU(2)$_{1}$ chiral edge
  modes of a critical spin liquid}},}\ }\href@noop {} {\bibfield  {journal}
  {\bibinfo  {journal} {\prb}\ }\textbf {\bibinfo {volume} {93}},\ \bibinfo
  {pages} {174414} (\bibinfo {year} {2016})}\BibitemShut {NoStop}%
\bibitem [{\citenamefont {{Zaletel}}\ and\ \citenamefont
  {{Mong}}(2012)}]{Zaletel2012a}%
  \BibitemOpen
  \bibfield  {author} {\bibinfo {author} {\bibfnamefont {M.~P.}\ \bibnamefont
  {{Zaletel}}}\ and\ \bibinfo {author} {\bibfnamefont {R.~S.~K.}\ \bibnamefont
  {{Mong}}},\ }\bibfield  {title} {\enquote {\bibinfo {title} {{Exact matrix
  product states for quantum Hall wave functions}},}\ }\href@noop {} {\bibfield
   {journal} {\bibinfo  {journal} {\prb}\ }\textbf {\bibinfo {volume} {86}},\
  \bibinfo {pages} {245305} (\bibinfo {year} {2012})}\BibitemShut {NoStop}%
\bibitem [{\citenamefont {Schmoll}\ and\ \citenamefont
  {Or{\'{u}}s}(2017)}]{Schmoll2017a}%
  \BibitemOpen
  \bibfield  {author} {\bibinfo {author} {\bibfnamefont {P.}~\bibnamefont
  {Schmoll}}\ and\ \bibinfo {author} {\bibfnamefont {R.}~\bibnamefont
  {Or{\'{u}}s}},\ }\bibfield  {title} {\enquote {\bibinfo {title} {{Kitaev
  honeycomb tensor networks: Exact unitary circuits and applications}},}\
  }\href@noop {} {\bibfield  {journal} {\bibinfo  {journal} {\prb}\ }\textbf
  {\bibinfo {volume} {95}},\ \bibinfo {pages} {045112} (\bibinfo {year}
  {2017})}\BibitemShut {NoStop}%
\bibitem [{\citenamefont {Nozaki}\ \emph {et~al.}(2012)\citenamefont {Nozaki},
  \citenamefont {Ryu},\ and\ \citenamefont {Takayanagi}}]{Nozaki2012}%
  \BibitemOpen
  \bibfield  {author} {\bibinfo {author} {\bibfnamefont {M.}~\bibnamefont
  {Nozaki}}, \bibinfo {author} {\bibfnamefont {S.}~\bibnamefont {Ryu}}, \ and\
  \bibinfo {author} {\bibfnamefont {T.}~\bibnamefont {Takayanagi}},\ }\bibfield
   {title} {\enquote {\bibinfo {title} {{Holographic geometry of entanglement
  renormalization in quantum field theories}},}\ }\href@noop {} {\bibfield
  {journal} {\bibinfo  {journal} {J. High Energy Phys.}\ }\textbf {\bibinfo
  {volume} {2012}},\ \bibinfo {pages} {193} (\bibinfo {year}
  {2012})}\BibitemShut {NoStop}%
\bibitem [{Note1()}]{Note1}%
  \BibitemOpen
  \bibinfo {note} {In the cMERA literature, a momentum cutoff $\Lambda $ is
  typically provided \cite {Haegeman2013,Wen2016}. With a finite cutoff, the UV
  state generated by a cMERA circuit approximates the ground state of the
  Hamiltonian up to $O(\protect \frac {1}{\Lambda })$ corrections. Here, we
  work in the continuum limit $\Lambda \rightarrow \infty $ to avoid this
  technical subtlety. In principle, these finite-$\Lambda $ corrections can be
  worked out explicitly.}\BibitemShut {Stop}%
\bibitem [{\citenamefont {Campbell}\ \emph {et~al.}(2011)\citenamefont
  {Campbell}, \citenamefont {Juzeliunas},\ and\ \citenamefont
  {Spielman}}]{Campbell2011}%
  \BibitemOpen
  \bibfield  {author} {\bibinfo {author} {\bibfnamefont {D.~L.}\ \bibnamefont
  {Campbell}}, \bibinfo {author} {\bibfnamefont {G.}~\bibnamefont
  {Juzeliunas}}, \ and\ \bibinfo {author} {\bibfnamefont {I.~B.}\ \bibnamefont
  {Spielman}},\ }\bibfield  {title} {\enquote {\bibinfo {title} {{Realistic
  Rashba and Dresselhaus spin-orbit coupling for neutral atoms}},}\ }\href@noop
  {} {\bibfield  {journal} {\bibinfo  {journal} {\pra}\ }\textbf {\bibinfo
  {volume} {84}},\ \bibinfo {pages} {025602} (\bibinfo {year}
  {2011})}\BibitemShut {NoStop}%
\bibitem [{\citenamefont {Campbell}\ and\ \citenamefont
  {Spielman}(2016)}]{Campbell2016}%
  \BibitemOpen
  \bibfield  {author} {\bibinfo {author} {\bibfnamefont {D.~L.}\ \bibnamefont
  {Campbell}}\ and\ \bibinfo {author} {\bibfnamefont {I.~B.}\ \bibnamefont
  {Spielman}},\ }\bibfield  {title} {\enquote {\bibinfo {title} {{Rashba
  realization: Raman with RF}},}\ }\href@noop {} {\bibfield  {journal}
  {\bibinfo  {journal} {New J. Phys.}\ }\textbf {\bibinfo {volume} {18}},\
  \bibinfo {pages} {033035} (\bibinfo {year} {2016})}\BibitemShut {NoStop}%
\bibitem [{\citenamefont {Huang}\ \emph {et~al.}(2016)\citenamefont {Huang},
  \citenamefont {Meng}, \citenamefont {Wang}, \citenamefont {Peng},
  \citenamefont {Zhang}, \citenamefont {Chen}, \citenamefont {Li},
  \citenamefont {Zhou},\ and\ \citenamefont {Zhang}}]{Huang2015a}%
  \BibitemOpen
  \bibfield  {author} {\bibinfo {author} {\bibfnamefont {L.}~\bibnamefont
  {Huang}}, \bibinfo {author} {\bibfnamefont {Z.}~\bibnamefont {Meng}},
  \bibinfo {author} {\bibfnamefont {P.}~\bibnamefont {Wang}}, \bibinfo {author}
  {\bibfnamefont {P.}~\bibnamefont {Peng}}, \bibinfo {author} {\bibfnamefont
  {S.-L.}\ \bibnamefont {Zhang}}, \bibinfo {author} {\bibfnamefont
  {L.}~\bibnamefont {Chen}}, \bibinfo {author} {\bibfnamefont {D.}~\bibnamefont
  {Li}}, \bibinfo {author} {\bibfnamefont {Q.}~\bibnamefont {Zhou}}, \ and\
  \bibinfo {author} {\bibfnamefont {J.}~\bibnamefont {Zhang}},\ }\bibfield
  {title} {\enquote {\bibinfo {title} {{Experimental realization of a
  two-dimensional synthetic spin-orbit coupling in ultracold Fermi gases}},}\
  }\href@noop {} {\bibfield  {journal} {\bibinfo  {journal} {Nat. Phys.}\
  }\textbf {\bibinfo {volume} {12}},\ \bibinfo {pages} {540} (\bibinfo {year}
  {2016})}\BibitemShut {NoStop}%
\bibitem [{\citenamefont {Galitski}\ and\ \citenamefont
  {Spielman}(2013)}]{Galitski2013}%
  \BibitemOpen
  \bibfield  {author} {\bibinfo {author} {\bibfnamefont {V.}~\bibnamefont
  {Galitski}}\ and\ \bibinfo {author} {\bibfnamefont {I.~B.}\ \bibnamefont
  {Spielman}},\ }\bibfield  {title} {\enquote {\bibinfo {title} {{Spin-orbit
  coupling in quantum gases}},}\ }\href@noop {} {\bibfield  {journal} {\bibinfo
   {journal} {\nat}\ }\textbf {\bibinfo {volume} {494}},\ \bibinfo {pages}
  {49--54} (\bibinfo {year} {2013})}\BibitemShut {NoStop}%
\bibitem [{\citenamefont {{Hauke}}\ \emph {et~al.}(2014)\citenamefont
  {{Hauke}}, \citenamefont {{Lewenstein}},\ and\ \citenamefont
  {{Eckardt}}}]{Hauke2014}%
  \BibitemOpen
  \bibfield  {author} {\bibinfo {author} {\bibfnamefont {P.}~\bibnamefont
  {{Hauke}}}, \bibinfo {author} {\bibfnamefont {M.}~\bibnamefont
  {{Lewenstein}}}, \ and\ \bibinfo {author} {\bibfnamefont {A.}~\bibnamefont
  {{Eckardt}}},\ }\bibfield  {title} {\enquote {\bibinfo {title} {{Tomography
  of Band Insulators from Quench Dynamics}},}\ }\href@noop {} {\bibfield
  {journal} {\bibinfo  {journal} {\prl}\ }\textbf {\bibinfo {volume} {113}},\
  \bibinfo {eid} {045303} (\bibinfo {year} {2014})}\BibitemShut {NoStop}%
\bibitem [{\citenamefont {{Alba}}\ \emph {et~al.}(2011)\citenamefont {{Alba}},
  \citenamefont {{Fernandez-Gonzalvo}}, \citenamefont {{Mur-Petit}},
  \citenamefont {{Pachos}},\ and\ \citenamefont
  {{Garcia-Ripoll}}}]{Alba:2011aa}%
  \BibitemOpen
  \bibfield  {author} {\bibinfo {author} {\bibfnamefont {E.}~\bibnamefont
  {{Alba}}}, \bibinfo {author} {\bibfnamefont {X.}~\bibnamefont
  {{Fernandez-Gonzalvo}}}, \bibinfo {author} {\bibfnamefont {J.}~\bibnamefont
  {{Mur-Petit}}}, \bibinfo {author} {\bibfnamefont {J.~K.}\ \bibnamefont
  {{Pachos}}}, \ and\ \bibinfo {author} {\bibfnamefont {J.~J.}\ \bibnamefont
  {{Garcia-Ripoll}}},\ }\bibfield  {title} {\enquote {\bibinfo {title} {{Seeing
  Topological Order in Time-of-Flight Measurements}},}\ }\href@noop {}
  {\bibfield  {journal} {\bibinfo  {journal} {\prl}\ }\textbf {\bibinfo
  {volume} {107}},\ \bibinfo {pages} {235301} (\bibinfo {year}
  {2011})}\BibitemShut {NoStop}%
\bibitem [{\citenamefont {{Fl{\"a}schner}}\ \emph {et~al.}(2016)\citenamefont
  {{Fl{\"a}schner}}, \citenamefont {{Rem}}, \citenamefont {{Tarnowski}},
  \citenamefont {{Vogel}}, \citenamefont {{L{\"u}hmann}}, \citenamefont
  {{Sengstock}},\ and\ \citenamefont {{Weitenberg}}}]{Flaschner:2016aa}%
  \BibitemOpen
  \bibfield  {author} {\bibinfo {author} {\bibfnamefont {N.}~\bibnamefont
  {{Fl{\"a}schner}}}, \bibinfo {author} {\bibfnamefont {B.~S.}\ \bibnamefont
  {{Rem}}}, \bibinfo {author} {\bibfnamefont {M.}~\bibnamefont {{Tarnowski}}},
  \bibinfo {author} {\bibfnamefont {D.}~\bibnamefont {{Vogel}}}, \bibinfo
  {author} {\bibfnamefont {D.-S.}\ \bibnamefont {{L{\"u}hmann}}}, \bibinfo
  {author} {\bibfnamefont {K.}~\bibnamefont {{Sengstock}}}, \ and\ \bibinfo
  {author} {\bibfnamefont {C.}~\bibnamefont {{Weitenberg}}},\ }\bibfield
  {title} {\enquote {\bibinfo {title} {{Experimental reconstruction of the
  Berry curvature in a Floquet Bloch band}},}\ }\href@noop {} {\bibfield
  {journal} {\bibinfo  {journal} {Science}\ }\textbf {\bibinfo {volume}
  {352}},\ \bibinfo {pages} {1091--1094} (\bibinfo {year} {2016})}\BibitemShut
  {NoStop}%
\bibitem [{\citenamefont {Anderson}\ \emph {et~al.}(2013)\citenamefont
  {Anderson}, \citenamefont {Spielman},\ and\ \citenamefont
  {Juzeliunas}}]{Anderson2013}%
  \BibitemOpen
  \bibfield  {author} {\bibinfo {author} {\bibfnamefont {B.~M.}\ \bibnamefont
  {Anderson}}, \bibinfo {author} {\bibfnamefont {I.~B.}\ \bibnamefont
  {Spielman}}, \ and\ \bibinfo {author} {\bibfnamefont {G.}~\bibnamefont
  {Juzeliunas}},\ }\bibfield  {title} {\enquote {\bibinfo {title}
  {{Magnetically generated spin-orbit coupling for ultracold atoms}},}\
  }\href@noop {} {\bibfield  {journal} {\bibinfo  {journal} {\prl}\ }\textbf
  {\bibinfo {volume} {111}},\ \bibinfo {pages} {125301} (\bibinfo {year}
  {2013})}\BibitemShut {NoStop}%
\bibitem [{\citenamefont {Grusdt}\ \emph {et~al.}(2017)\citenamefont {Grusdt},
  \citenamefont {Li}, \citenamefont {Bloch},\ and\ \citenamefont
  {Demler}}]{Grusdt2017}%
  \BibitemOpen
  \bibfield  {author} {\bibinfo {author} {\bibfnamefont {F.}~\bibnamefont
  {Grusdt}}, \bibinfo {author} {\bibfnamefont {T.}~\bibnamefont {Li}}, \bibinfo
  {author} {\bibfnamefont {I.}~\bibnamefont {Bloch}}, \ and\ \bibinfo {author}
  {\bibfnamefont {E.}~\bibnamefont {Demler}},\ }\bibfield  {title} {\enquote
  {\bibinfo {title} {{Tunable spin-orbit coupling for ultracold atoms in
  two-dimensional optical lattices}},}\ }\href@noop {} {\bibfield  {journal}
  {\bibinfo  {journal} {\pra}\ }\textbf {\bibinfo {volume} {95}},\ \bibinfo
  {pages} {063617} (\bibinfo {year} {2017})}\BibitemShut {NoStop}%
\bibitem [{\citenamefont {Huckans}\ \emph {et~al.}(2009)\citenamefont
  {Huckans}, \citenamefont {Spielman}, \citenamefont {Tolra}, \citenamefont
  {Phillips},\ and\ \citenamefont {Porto}}]{Huckans2009}%
  \BibitemOpen
  \bibfield  {author} {\bibinfo {author} {\bibfnamefont {J.~H.}\ \bibnamefont
  {Huckans}}, \bibinfo {author} {\bibfnamefont {I.~B.}\ \bibnamefont
  {Spielman}}, \bibinfo {author} {\bibfnamefont {B.~L.}\ \bibnamefont {Tolra}},
  \bibinfo {author} {\bibfnamefont {W.~D.}\ \bibnamefont {Phillips}}, \ and\
  \bibinfo {author} {\bibfnamefont {J.~V.}\ \bibnamefont {Porto}},\ }\bibfield
  {title} {\enquote {\bibinfo {title} {{Quantum and classical dynamics of a
  Bose-Einstein condensate in a large-period optical lattice}},}\ }\href@noop
  {} {\bibfield  {journal} {\bibinfo  {journal} {\pra}\ }\textbf {\bibinfo
  {volume} {80}},\ \bibinfo {pages} {043609} (\bibinfo {year}
  {2009})}\BibitemShut {NoStop}%
\bibitem [{\citenamefont {Al-Assam}\ \emph {et~al.}(2010)\citenamefont
  {Al-Assam}, \citenamefont {Williams},\ and\ \citenamefont
  {Foot}}]{Al-Assam2010}%
  \BibitemOpen
  \bibfield  {author} {\bibinfo {author} {\bibfnamefont {S.}~\bibnamefont
  {Al-Assam}}, \bibinfo {author} {\bibfnamefont {R.~A.}\ \bibnamefont
  {Williams}}, \ and\ \bibinfo {author} {\bibfnamefont {C.~J.}\ \bibnamefont
  {Foot}},\ }\bibfield  {title} {\enquote {\bibinfo {title} {{Ultracold atoms
  in an optical lattice with dynamically variable periodicity}},}\ }\href@noop
  {} {\bibfield  {journal} {\bibinfo  {journal} {\pra}\ }\textbf {\bibinfo
  {volume} {82}},\ \bibinfo {pages} {021604} (\bibinfo {year}
  {2010})}\BibitemShut {NoStop}%
\bibitem [{\citenamefont {Fukuhara}\ \emph {et~al.}(2013)\citenamefont
  {Fukuhara}, \citenamefont {Kantian}, \citenamefont {Endres}, \citenamefont
  {Cheneau}, \citenamefont {Schau{\ss}}, \citenamefont {Hild}, \citenamefont
  {Bellem}, \citenamefont {Schollw{\"o}ck}, \citenamefont {Giamarchi},
  \citenamefont {Gross}, \citenamefont {Bloch},\ and\ \citenamefont
  {Kuhr}}]{Fukuhara:2013aa}%
  \BibitemOpen
  \bibfield  {author} {\bibinfo {author} {\bibfnamefont {T.}~\bibnamefont
  {Fukuhara}}, \bibinfo {author} {\bibfnamefont {A.}~\bibnamefont {Kantian}},
  \bibinfo {author} {\bibfnamefont {M.}~\bibnamefont {Endres}}, \bibinfo
  {author} {\bibfnamefont {M.}~\bibnamefont {Cheneau}}, \bibinfo {author}
  {\bibfnamefont {P.}~\bibnamefont {Schau{\ss}}}, \bibinfo {author}
  {\bibfnamefont {S.}~\bibnamefont {Hild}}, \bibinfo {author} {\bibfnamefont
  {D.}~\bibnamefont {Bellem}}, \bibinfo {author} {\bibfnamefont
  {U.}~\bibnamefont {Schollw{\"o}ck}}, \bibinfo {author} {\bibfnamefont
  {T.}~\bibnamefont {Giamarchi}}, \bibinfo {author} {\bibfnamefont
  {C.}~\bibnamefont {Gross}}, \bibinfo {author} {\bibfnamefont
  {I.}~\bibnamefont {Bloch}}, \ and\ \bibinfo {author} {\bibfnamefont
  {S.}~\bibnamefont {Kuhr}},\ }\bibfield  {title} {\enquote {\bibinfo {title}
  {Quantum dynamics of a mobile spin impurity},}\ }\href
  {http://dx.doi.org/10.1038/nphys2561} {\bibfield  {journal} {\bibinfo
  {journal} {Nat. Phys.}\ }\textbf {\bibinfo {volume} {9}},\ \bibinfo {pages}
  {235 EP --} (\bibinfo {year} {2013})}\BibitemShut {NoStop}%
\bibitem [{\citenamefont {Nogrette}\ \emph {et~al.}(2014)\citenamefont
  {Nogrette}, \citenamefont {Labuhn}, \citenamefont {Ravets}, \citenamefont
  {Barredo}, \citenamefont {B\'eguin}, \citenamefont {Vernier}, \citenamefont
  {Lahaye},\ and\ \citenamefont {Browaeys}}]{Nogrette2014}%
  \BibitemOpen
  \bibfield  {author} {\bibinfo {author} {\bibfnamefont {F.}~\bibnamefont
  {Nogrette}}, \bibinfo {author} {\bibfnamefont {H.}~\bibnamefont {Labuhn}},
  \bibinfo {author} {\bibfnamefont {S.}~\bibnamefont {Ravets}}, \bibinfo
  {author} {\bibfnamefont {D.}~\bibnamefont {Barredo}}, \bibinfo {author}
  {\bibfnamefont {L.}~\bibnamefont {B\'eguin}}, \bibinfo {author}
  {\bibfnamefont {A.}~\bibnamefont {Vernier}}, \bibinfo {author} {\bibfnamefont
  {T.}~\bibnamefont {Lahaye}}, \ and\ \bibinfo {author} {\bibfnamefont
  {A.}~\bibnamefont {Browaeys}},\ }\bibfield  {title} {\enquote {\bibinfo
  {title} {Single-atom trapping in holographic 2d arrays of microtraps with
  arbitrary geometries},}\ }\href {\doibase 10.1103/PhysRevX.4.021034}
  {\bibfield  {journal} {\bibinfo  {journal} {Phys. Rev. X}\ }\textbf {\bibinfo
  {volume} {4}},\ \bibinfo {pages} {021034} (\bibinfo {year}
  {2014})}\BibitemShut {NoStop}%
\bibitem [{\citenamefont {Zhang}\ \emph {et~al.}(2015)\citenamefont {Zhang},
  \citenamefont {Zhou}, \citenamefont {Chen}, \citenamefont {Gao},
  \citenamefont {Han}, \citenamefont {Yao}, \citenamefont {Xu}, \citenamefont
  {Li}, \citenamefont {Xu}, \citenamefont {Jiang}, \citenamefont {Bi},
  \citenamefont {Ma},\ and\ \citenamefont {Xu}}]{Zhang2015}%
  \BibitemOpen
  \bibfield  {author} {\bibinfo {author} {\bibfnamefont {X.}~\bibnamefont
  {Zhang}}, \bibinfo {author} {\bibfnamefont {M.}~\bibnamefont {Zhou}},
  \bibinfo {author} {\bibfnamefont {N.}~\bibnamefont {Chen}}, \bibinfo {author}
  {\bibfnamefont {Q.}~\bibnamefont {Gao}}, \bibinfo {author} {\bibfnamefont
  {C.}~\bibnamefont {Han}}, \bibinfo {author} {\bibfnamefont {Y.}~\bibnamefont
  {Yao}}, \bibinfo {author} {\bibfnamefont {P.}~\bibnamefont {Xu}}, \bibinfo
  {author} {\bibfnamefont {S.}~\bibnamefont {Li}}, \bibinfo {author}
  {\bibfnamefont {Y.}~\bibnamefont {Xu}}, \bibinfo {author} {\bibfnamefont
  {Y.}~\bibnamefont {Jiang}}, \bibinfo {author} {\bibfnamefont
  {Z.}~\bibnamefont {Bi}}, \bibinfo {author} {\bibfnamefont {L.}~\bibnamefont
  {Ma}}, \ and\ \bibinfo {author} {\bibfnamefont {X.}~\bibnamefont {Xu}},\
  }\bibfield  {title} {\enquote {\bibinfo {title} {{Study on the
  clock-transition spectrum of cold 171 Yb ytterbium atoms}},}\ }\href@noop {}
  {\bibfield  {journal} {\bibinfo  {journal} {Laser Phys. Lett.}\ }\textbf
  {\bibinfo {volume} {12}},\ \bibinfo {pages} {25501} (\bibinfo {year}
  {2015})}\BibitemShut {NoStop}%
\bibitem [{\citenamefont {Kohno}\ \emph {et~al.}(2009)\citenamefont {Kohno},
  \citenamefont {Yasuda}, \citenamefont {Hosaka}, \citenamefont {Inaba},
  \citenamefont {Nakajima},\ and\ \citenamefont {Hong}}]{Kohno2009}%
  \BibitemOpen
  \bibfield  {author} {\bibinfo {author} {\bibfnamefont {T.}~\bibnamefont
  {Kohno}}, \bibinfo {author} {\bibfnamefont {M.}~\bibnamefont {Yasuda}},
  \bibinfo {author} {\bibfnamefont {K.}~\bibnamefont {Hosaka}}, \bibinfo
  {author} {\bibfnamefont {H.}~\bibnamefont {Inaba}}, \bibinfo {author}
  {\bibfnamefont {Y.}~\bibnamefont {Nakajima}}, \ and\ \bibinfo {author}
  {\bibfnamefont {F.-L.}\ \bibnamefont {Hong}},\ }\bibfield  {title} {\enquote
  {\bibinfo {title} {{One-Dimensional Optical Lattice Clock with a Fermionic
  171 Yb Isotope}},}\ }\href@noop {} {\bibfield  {journal} {\bibinfo  {journal}
  {Appl. Phys. Express}\ }\textbf {\bibinfo {volume} {2}},\ \bibinfo {pages}
  {072501} (\bibinfo {year} {2009})}\BibitemShut {NoStop}%
\bibitem [{\citenamefont {Lemke}\ \emph {et~al.}(2009)\citenamefont {Lemke},
  \citenamefont {Ludlow}, \citenamefont {Barber}, \citenamefont {Fortier},
  \citenamefont {Diddams}, \citenamefont {Jiang}, \citenamefont {Jefferts},
  \citenamefont {Heavner}, \citenamefont {Parker},\ and\ \citenamefont
  {Oates}}]{Lemke2009}%
  \BibitemOpen
  \bibfield  {author} {\bibinfo {author} {\bibfnamefont {N.~D.}\ \bibnamefont
  {Lemke}}, \bibinfo {author} {\bibfnamefont {A.~D.}\ \bibnamefont {Ludlow}},
  \bibinfo {author} {\bibfnamefont {Z.~W.}\ \bibnamefont {Barber}}, \bibinfo
  {author} {\bibfnamefont {T.~M.}\ \bibnamefont {Fortier}}, \bibinfo {author}
  {\bibfnamefont {S.~A.}\ \bibnamefont {Diddams}}, \bibinfo {author}
  {\bibfnamefont {Y.}~\bibnamefont {Jiang}}, \bibinfo {author} {\bibfnamefont
  {S.~R.}\ \bibnamefont {Jefferts}}, \bibinfo {author} {\bibfnamefont {T.~P.}\
  \bibnamefont {Heavner}}, \bibinfo {author} {\bibfnamefont {T.~E.}\
  \bibnamefont {Parker}}, \ and\ \bibinfo {author} {\bibfnamefont {C.~W.}\
  \bibnamefont {Oates}},\ }\bibfield  {title} {\enquote {\bibinfo {title}
  {{Spin-1/2 optical lattice clock}},}\ }\href@noop {} {\bibfield  {journal}
  {\bibinfo  {journal} {\prl}\ }\textbf {\bibinfo {volume} {103}},\ \bibinfo
  {pages} {063001} (\bibinfo {year} {2009})}\BibitemShut {NoStop}%
\bibitem [{\citenamefont {Park}\ \emph {et~al.}(2013)\citenamefont {Park},
  \citenamefont {Yu}, \citenamefont {Lee}, \citenamefont {Park}, \citenamefont
  {Kim}, \citenamefont {Lee}, \citenamefont {Cho}, \citenamefont {Yoon},
  \citenamefont {Mun}, \citenamefont {Park}, \citenamefont {Kwon},\ and\
  \citenamefont {Lee}}]{Park2013}%
  \BibitemOpen
  \bibfield  {author} {\bibinfo {author} {\bibfnamefont {C.~Y.}\ \bibnamefont
  {Park}}, \bibinfo {author} {\bibfnamefont {D.-H.}\ \bibnamefont {Yu}},
  \bibinfo {author} {\bibfnamefont {W.-K.}\ \bibnamefont {Lee}}, \bibinfo
  {author} {\bibfnamefont {S.~E.}\ \bibnamefont {Park}}, \bibinfo {author}
  {\bibfnamefont {E.~B.}\ \bibnamefont {Kim}}, \bibinfo {author} {\bibfnamefont
  {S.~K.}\ \bibnamefont {Lee}}, \bibinfo {author} {\bibfnamefont {J.~W.}\
  \bibnamefont {Cho}}, \bibinfo {author} {\bibfnamefont {T.~H.}\ \bibnamefont
  {Yoon}}, \bibinfo {author} {\bibfnamefont {J.}~\bibnamefont {Mun}}, \bibinfo
  {author} {\bibfnamefont {S.~J.}\ \bibnamefont {Park}}, \bibinfo {author}
  {\bibfnamefont {T.~Y.}\ \bibnamefont {Kwon}}, \ and\ \bibinfo {author}
  {\bibfnamefont {S.-B.}\ \bibnamefont {Lee}},\ }\bibfield  {title} {\enquote
  {\bibinfo {title} {{Absolute frequency measurement of transition of
  $^{1}$S$_{0}$(F=1/2)-$^3$P$_{0}$(F=1/2) $^{171}$Yb atoms in a one-dimensional
  optical lattice at KRISS}},}\ }\href@noop {} {\bibfield  {journal} {\bibinfo
  {journal} {Metrologia}\ }\textbf {\bibinfo {volume} {50}},\ \bibinfo {pages}
  {119--128} (\bibinfo {year} {2013})}\BibitemShut {NoStop}%
\bibitem [{\citenamefont {Yamaguchi}(2008)}]{Yamaguchi2008}%
  \BibitemOpen
  \bibfield  {author} {\bibinfo {author} {\bibfnamefont {A.}~\bibnamefont
  {Yamaguchi}},\ }\emph {\bibinfo {title} {Metastable State of Ultracold and
  Quantum Degenerate Ytterbium Atoms: High-Resolution Spectroscopy and Cold
  Collisions}},\ \href@noop {} {Ph.D. thesis},\ \bibinfo  {school} {Kyoto
  University} (\bibinfo {year} {2008})\BibitemShut {NoStop}%
\bibitem [{\citenamefont {Busch}\ and\ \citenamefont
  {Penson}(1987)}]{Busch1987}%
  \BibitemOpen
  \bibfield  {author} {\bibinfo {author} {\bibfnamefont {U.}~\bibnamefont
  {Busch}}\ and\ \bibinfo {author} {\bibfnamefont {K.~A.}\ \bibnamefont
  {Penson}},\ }\bibfield  {title} {\enquote {\bibinfo {title} {Tight-binding
  electrons on open chains: Density distribution and correlations},}\
  }\href@noop {} {\bibfield  {journal} {\bibinfo  {journal} {\prb}\ }\textbf
  {\bibinfo {volume} {36}},\ \bibinfo {pages} {9271--9274} (\bibinfo {year}
  {1987})}\BibitemShut {NoStop}%
\end{thebibliography}
\end{document}